\DocumentMetadata{lang={en-GB},tagging=on}
\documentclass[a4paper]{article}
\usepackage[LGR,T2A,T1]{fontenc}
\usepackage[final]{hyperref}
\usepackage[greek,russian,ngerman,french,italian,british]{babel}
\usepackage{amsmath,amssymb,amsthm,orcidlink,cleveref,tikz-cd,mathtools,mleftright}
\usetikzlibrary{babel}
\usepackage{CJKutf8}
\usepackage{keytheorems}
\usepackage[final]{microtype}
\newcommand\langlands[1]{{}^{\mathrm L}{#1}}
\renewcommand\Re{\operatorname{Re}}
\newcommand\piit\pi
\renewcommand\pi{\text{\textrm{\greektext{p}}}}
\renewcommand\Sigma\varSigma

\newtheorem{theorem}{Theorem}
\newkeytheorem{postulate}
\theoremstyle{definition}
\newtheorem{definition}{Definition}
\newtheorem{constraint}{Constraint}
\begin{document}
\title{{Generalised Symmetries and Swampland-Type Constraints from Charge Quantisation via Rational Homotopy Theory}}
\author[pdfauthor={{Luigi Alfonsi, 金炯錄, William G. A. Luciani}}]
{Luigi Alfonsi\textsuperscript{\orcidlink{0000-0001-5231-2354}}\footnote{Centre for Mathematics and Theoretical Physics Research, Department of Physics, Astronomy and Mathematics, University of Hertfordshire, Hatfield, Hertfordshire\ AL10~9AB, United Kingdom}
\and Hyungrok Kim (\begin{CJK*}{UTF8}{bsmi}金炯錄\end{CJK*})\textsuperscript{\orcidlink{0000-0001-7909-4510}}\footnotemark[1]\and William G. A. Luciani\textsuperscript{\orcidlink{0009-0007-7443-9955}}\footnotemark[1]}
\maketitle
\begin{abstract}
Sati and Schreiber \cite{Sati:2024klw,Sati:2025vjw} have proposed that charge quantisation in quantum field theory and string theory is governed by a homotopy type \(\mathcal A\).
We provide a refinement of this postulate, incorporating other currents including matter, connecting it to adjustments in higher gauge theory and providing a prescription for determining \(\mathcal A\), and show that, while the homotopy groups of \(\mathcal A\) classify the possible brane charges, the homology groups of \(\mathcal A\) classify the invertible higher-form symmetries.
Furthermore, we show that the charge-quantisation postulate implies a number of non-trivial constraints on quantum field theories similar to those implied by swampland conjectures; in particular, it rules out noncompact gauge groups and one-form field strengths that form a non-nilpotent Lie algebra.
Finally, we argue that for theories of quantum gravity the space \(\mathcal A\) must be contractible, in accordance with the swampland conjectures on the absence of global generalised symmetries and the completeness of the spectrum of charges, and explain how this explicitly arises in the case of Type~I string theory.
\end{abstract}
\tableofcontents
\section{Introduction}
Symmetries, their associated charges, and defects that carry these charges are crucial aspects of a physical system.
In particular, generalised symmetries \cite{Gaiotto:2014kfa}, reviewed in \cite{Cordova:2022ruw,Brennan:2023mmt,Gomes:2023ahz,Bhardwaj:2023kri,Luo:2023ive,Iqbal:2024pee},
arise in most physical theories of interest, including Maxwell theory, Yang--Mills theory, and string theory,
and have broadened and enriched our understanding of symmetries in quantum field theories.

Recently, Sati and Schreiber have argued \cite{Sati:2024klw,Sati:2025vjw} that charges of branes and other solitonic objects can be understood in terms of the homotopy groups of a homotopy type \(\mathcal A\), which we will call the \emph{classifying space} of the theory.
The data of \(\mathcal A\) cannot be read off simply from the local equations of motion but constitutes additional data in the specification of the theory, similar to how \(\operatorname{SU}(2)\)- and \(\operatorname{SO}(3)\)-valued
Yang--Mills theories have identical local equations of motion but only differ in global aspects (\(\operatorname{SU}(2)\)-bundles versus \(\operatorname{SO}(3)\)-bundles) and in the charge quantisation laws.
Examples of such classifying spaces of theories have been given for Maxwell theory \cite{Sati:2024klw}, M-theory \cite{Fiorenza:2019ain,Fiorenza:2019usl,Sati:2020cml}, the M5-brane worldvolume theory \cite{Giotopoulos:2024sit}, and for anyonic systems exhibiting the fractional quantum Hall effect \cite{Sati:2025vjw,Sati:2025iiw,Sati:2025gxj,Sati:2025ras,Sati_2025}. In particular, brane charges on flat spacetime correspond to homotopy groups \(\pi_\bullet(\mathcal A)\) of the classifying space.

Given the close relationship between symmetries and corresponding conserved charges given by Noether's theorem,
it is therefore a natural question as to whether and how generalised symmetries are encoded in the classifying space.
In this paper, we argue that invertible higher-form symmetries correspond to the Pontryagin duals of the \emph{homology groups} \(\operatorname H_\bullet(\mathcal A)\) of the classifying space. We construct classifying spaces of Maxwell theory, Yang--Mills theory, and \(\mathcal N=(1,0)\) string theories and show that this proposal recovers expected electric and magnetic higher-form symmetries, including the discrete centre one-form symmetries of Yang--Mills theory that play a crucial role in confinement \cite{Polyakov:1975rs,Polyakov:1976fu,tHooft:1977nqb}.

As a byproduct, we argue that the charge-quantisation postulate of Sati and Schreiber imply significant
constraints on the consistency of quantum field theories, similar in flavour to some of the swampland conjectures (reviewed in \cite{Brennan:2017rbf,Palti:2019pca,Grana:2021zvf,vanBeest:2021lhn,Agmon:2022thq,Lehnert:2025izp}).
In particular, we show that the charge-quantisation proposal rules out noncompact gauge groups (and, consequently, noncompact global symmetry groups that are free of 't~Hooft anomalies); this is also implied by the swampland completeness conjecture \cite{vanBeest:2021lhn,Tadros:2022rkd}. Similarly, the charge-quantisation postulate also rules out certain Bianchi identities involving one-form field strengths; in particular, their Lie algebra must be nilpotent.

\paragraph{Related work and future directions.}
Symmetry topological field theories (SymTFTs) programme \cite{Apruzzi:2021nmk,Bhardwaj:2023ayw} (reviewed in \cite{Schafer-Nameki:2023jdn}) provides a characterisation and classification of higher-form symmetries of a \(d\)-dimensional field theory in terms of a \((d+1)\)-dimensional topological field theory.
These are more general than the generalised symmetries we consider insofar as we only obtain \emph{invertible} generalised symmetries.
In this paper, we do not attempt to make contact with SymTFTs, although we note that \(L_\infty\)-algebras, which are a key ingredient in the present discussion, also appear in constructions of SymTFTs \cite{Pulmann:2019vrw,Borsten:2025pvq}.

Classifying spaces of (higher) bundles have recently appeared in connection with generalised symmetries \cite{Perez-Lona:2025xxm}, but here it is the (classifying spaces of) higher \emph{automorphisms} of bundles that detect the centre symmetry, rather than the classifying spaces of bundles themselves (cf.\ \cite{Sati:2024gqu}). It would be interesting to connect the results presented here with the automorphism-based approach.

\paragraph{Organisation of this paper.}
We start by reviewing, in \cref{sec:review}, the relevant background in \(L_\infty\)-algebras and rational homotopy theory needed for the rest of the discussion, including establishing conventions and notation.
In \cref{sec:quantisation}, we review Sati and Schreiber's charge-quantisation postulate and provide a refined version of the postulate (\cref{post:charge-quantisation}), including a prescription for how the classifying space \(\mathcal A\) is to be obtained in the general case.
In \cref{sec:swampland}, we explain how the charge-quantisation postulate rules out various classes of quantum field theories, including theories with noncompact gauge groups or with non-nilpotent algebras of one-forms.
Finally in \cref{sec:nlsm,sec:abelian,sec:yang-mills,sec:adjusted}, we examine how our construction reproduces known higher-form symmetries, such as the electric and magnetic higher-form symmetries of Abelian \(p\)-form gauge theory and the electric and magnetic centre symmetries of Yang--Mills theory, and construct exotic higher-form symmetries corresponding to torsion components of higher homotopy groups.

\section{Review of homotopy theory}\label{sec:review}
Let us briefly review elements of homotopy algebras and rational homotopy theory that we will need in the following. Proofs are not given; we have included citations to reviews and books, where the reader can find more detailed treatments.
We work over a general field \(\mathbb K\) of characteristic zero since, although the de~Rham algebra is naturally defined over the field \(\mathbb R\) of real numbers, we will later want to work over the field \(\mathbb Q\) of rational numbers instead.

\subsection{Differential graded-commutative algebras}
We first start with an abstraction of the de~Rham algebra of differential forms and the exterior derivative, the differential graded-commutative algebra (reviewed in \cite{hess}).
\begin{definition} 
A \emph{differential graded-commutative algebra} (or \emph{dgca}) \(A=\bigoplus_{i\in\mathbb Z}A^i\) over \(\mathbb K\) is a unital graded algebra over \(\mathbb K\) obeying the graded-commutativity condition, i.e.\
\begin{equation}
    xy=(-)^{|x||y|}yx
\end{equation}
for elements \(x\) and \(y\) of degrees \(|x|\) and \(|y|\), equipped with a linear map \(\mathrm d\colon A\to A\) of degree \(1\) that squares to zero (i.e.\ \(\mathrm d^2=0\)) and obeys the graded Leibniz rule
\begin{equation}
    \mathrm d(xy) = (\mathrm dx)y+(-1)^{|x|}x(\mathrm dy).
\end{equation}
A \emph{morphism} between two dgcas is an algebra morphism preserving degrees that is also a cochain map with respect to the differential. A \emph{quasi-isomorphism} between two dgcas is a dgca morphism \(\phi\colon A\to B\) that induces an isomorphism on the cohomologies \(\operatorname H^\bullet(A)\to\operatorname H^\bullet(B)\).
\end{definition}

We wish to look at those differential graded-commutative algebras that can be regarded as the de~Rham algebra of a connected space with finite Betti numbers. Consider the category \(\operatorname{dgcAlg}_{>0}\) of differential graded-commutative algebras \(A\) concentrated in non-negative degree that are degreewise finite, whose degree-zero cohomology \(\operatorname H^0(A) = \mathbb K\) is simply the underlying field. The category \(\operatorname{dgcAlg}_{>0}\) can be promoted to an \(\infty\)-category; more precisely, there exists a model-category structure on \(\operatorname{dgcAlg}_{>0}\) in which the weak equivalences are the quasi-isomorphisms.

Because of this \(\infty\)-category structure, it is natural to look for invariants of quasi-isomorphism classes. One obvious invariant is the cohomology: any dgca \(A\) is, in particular, a cochain complex, and one can compute its cohomology \(\operatorname H^\bullet(A)\). We can, however, do much better: there exists a canonical representative of a dgca belonging to \(\operatorname{dgcAlg}_{>0}\). We write the definition for $\mathbb K = \mathbb Q$ as it will be the case we'll mostly use in the following.
\begin{definition}[\cite{hess}]
    A dgca \(A\) in \(\operatorname{dgcAlg}_{>0}\) is a \emph{minimal connected degreewise-finite Sullivan algebra} if its underlying algebra structure is of the form
    \begin{equation}
        A = \mathbb Q[x_1,x_2,\dotsc]
    \end{equation}
    in countably many generators \(x_i\), such that the derivative \(\mathrm dx_i\) only involves \(x_1,\dotsc,x_{i-1}\), and \(i<j\) implies \(|x_i|\le|x_j|\).
\end{definition}

\begin{theorem}[{{\cite[Theorem~2.24]{felix}}}]
    For any dgca \(A\) in \(\operatorname{dgcAlg}_{>0}\), the quasi-isomorphism equivalence class contains exactly one minimal connected degreewise-finite Sullivan algebra, which we call the \emph{Sullivan model} \(A_\mathrm{Sull}\).
\end{theorem}
In particular, the number of generators in degree \(i\) of the Sullivan model \(A_\mathrm{Sull}\) is an invariant of quasi-isomorphisms.

\subsection{\texorpdfstring{\(L_\infty\)}{L∞}-algebras}
\(L_\infty\)-algebras, sometimes called (strong) homotopy Lie algebras, are a generalisation of graded Lie algebras in which the Jacobi identity (but \emph{not} the antisymmetry of the Lie bracket) is relaxed up to homotopy (see \cite{Kraft:2022efy,Jurco:2018sby,Lada:1992wc} for reviews).
\begin{definition}
    An \(L_\infty\)-algebra \((\mathfrak g,(\mu_i)_{i=1}^\infty)\) is a \(\mathbb Z\)-graded vector space \(\mathfrak g=\bigoplus_{i\in\mathbb Z}\mathfrak g^i\) together with totally graded-antisymmetric maps of arity \(i\)
    \begin{equation}
        \mu_i\colon\overbrace{\mathfrak g\otimes\dotso\otimes\mathfrak g}^i\to\mathfrak g
    \end{equation}
    of degree \(d-2i\) that obey the \emph{homotopy Jacobi identities}
    \begin{equation}\label{eq:homotopy-jacobi}
        \sum_{\substack{i+j=k\\\sigma\in\operatorname{Sym}(k)}}
        \frac{(-1)^{ij}}{i!j!}\chi(\sigma)\mu_{i+1}(\mu_i(x_{\sigma(1)},\dotsc,x_{\sigma(i)}),x_{\sigma(i+1)},\dotsc,x_{\sigma(k)}) = 0
    \end{equation}
    for any homogeneous elements \(x_1,\dotsc,x_k\in\mathfrak g\) and positive integer \(k\in\mathbb Z^+\), where \(\chi(\sigma)\in\{\pm1\}\) is the Koszul sign defined by
    \begin{equation}
        \chi(\sigma)x_1\wedge\dotsb\wedge x_k\coloneqq x_{\sigma(1)}\wedge\dotsb\wedge x_{\sigma(k)}.
    \end{equation}
\end{definition}
In particular, an \(L_\infty\)-algebra concentrated in degree zero is a Lie algebra since \(\mu_i=0\) except for the Lie bracket \(\mu_2\), and the homotopy Jacobi identity \eqref{eq:homotopy-jacobi} then reduces to the ordinary Jacobi identity.

\(L_\infty\)-algebras are more conveniently thought of in terms of their dual dgcas, the \emph{Chevalley--Eilenberg algebra}.
\begin{definition}
Given an \(L_\infty\)-algebra \(\mathfrak g\) that is degreewise finite\footnote{i.e.\ \(\dim\mathfrak g^i <\infty\) for every \(i\in\mathbb Z\)}, its \emph{Chevalley--Eilenberg algebra} \(\operatorname{CE}(\mathfrak g)\) is the quasifree dgca
\begin{equation}
    \operatorname{CE}(\mathfrak g) = \bigodot\mathfrak g^*[-1]
\end{equation}
with differential given by
\begin{equation}
    \mathrm d\mathtt t^i = \sum_{n=1}^\infty -\frac1{n!}f^i_{j_1\dotso j_n}\mathtt t^{j_1}\dotsm\mathtt t^{j_n},
\end{equation}
where \(f^i_{j_1\dotso j_n}\) are the structure constants of \(\mu_n\).
A morphism of (degreewise finite) \(L_\infty\)-algebras \(\mathfrak g\to\mathfrak h\) is the same as a morphism of differential graded-commutative algebras \(\operatorname{CE}(\mathfrak h)\to\operatorname{CE}(\mathfrak g)\).
\end{definition}

We thus can obtain a quasifree dgca starting from a degreewise finite \(L_\infty\)-algebra and by Koszul duality the converse is also true.
This applies in particular to Sullivan models: if \(A\) is a (not necessarily quasifree) dgca belonging to \(\operatorname{dgcAlg}_{>0}\), then its Sullivan model \(A_\mathrm{Sull}\) is the Chevalley--Eilenberg algebra of an \(L_\infty\)-algebra that is degreewise finite and concentrated in non-positive degrees.

\subsection{Rational homotopy theory}\label{ssec:rational-homotopy-theory}
Homotopy theory aims to classify spaces up to weak homotopy equivalence,
but in general this turns out to be a very difficult task, in large part due to the torsion of homotopy and cohomology groups.
Rational homotopy theory \cite{quillen,Sullivan:1977pdi} (reviewed in \cite{hess,griffiths,felix1,felix2,felix,Voronov:2024lon}) aims at a coarser classification
where difficulties related to torsion are absent but which still suffices to compute the ranks of homotopy and cohomology groups.

The central objects studied are homotopy types which are a very coarse classification of topological spaces that nevertheless preserve key invariants such as homotopy and homology groups. A \emph{weak homotopy equivalence} \(f\colon X\to Y\) is a continuous map such that, for any \(x\in X\), the induced map \(\pi_i(X,x)\to\pi_i(Y,f(x))\) is an isomorphism.
Weak homotopy equivalences generate an equivalence relation.
A \emph{homotopy type} is an equivalence class of topological spaces under weak homotopy equivalences. In particular, homotopy groups \(\pi_\bullet(X,x)\) and (co)homology groups \(\operatorname H_\bullet(X)\), \(\operatorname H^\bullet(X)\) are invariants of homotopy type; in particular, the homotopy type knows about the torsion of homotopy and (co)homology groups.

A rational homotopy type is an even coarser classification of topological spaces than the homotopy type, which only remembers information about the torsion-free parts of homotopy and cohomology groups. That is, a \emph{rational homotopy equivalence} \(f\colon X\to Y\) between two path-connected spaces \(X\) and \(Y\) with nilpotent fundamental groups is a continuous map such that, for any \(x\in X\), the induced map \(\pi_i(X,x)\otimes_{\mathbb Z}\mathbb Q\to\pi_i(Y,f(x))\otimes_{\mathbb Z}\mathbb Q\) is an isomorphism. (This presents difficulties for \(i=1\) where the homotopy group is not Abelian; in this case, one constructs \(\pi_1(-)\otimes\mathbb Q\) by replacing \(\pi_1(-)\) by a tower of nilpotent groups for which one can define the tensor product with \(\mathbb Q\), see \cite[§13.2]{griffiths}.) Rational homotopy equivalences generate an equivalence relation, whose equivalence classes are called \emph{rational homotopy types}.

The major result of rational homotopy theory is that rational homotopy types are adequately captured by an analogue of the de~Rham complex of differential forms, but defined over the rational numbers rather than the reals. More precisely, this works for a class of `nice' spaces, namely the nilpotent spaces with finite Betti numbers\footnote{For a discussion of the complications beyond this case, see\ \cite{ivanov}.}, which are defined as follows.

In any path-connected space \(X\), the fundamental group \(\pi_1(X)\) canonically acts on \(\pi_k(X)\) as follows \cite[p.~30--31]{felix}. Given a basepoint \(\bullet\in\mathbb S^k\) and an element \(\alpha\in\pi_1(X)\) and \(\xi\in\pi_k(X)\), then consider the following diagram.
\begin{equation}
    \begin{tikzcd}[ampersand replacement=\&]
        \mathbb S^n\times[0,1] \ar[dashed]{dr}{F} \& \bullet\times[0,1] \lar \ar{d}{\alpha} \\
        \mathbb S^k\times\{0\} \uar \ar{r}[below]{\xi} \& X
    \end{tikzcd}
\end{equation}
The existence of a map \(F\) is guaranteed by by the cofibration property of the inclusion \(\bullet\hookrightarrow \mathbb S^k\). Then \(\alpha\cdot\xi\) is defined to be the homotopy class of \(F(-,1)\), which is unique.
\def\foo{\cite[Def.~2.32]{felix}}
\begin{definition}[\foo]
    A \emph{nilpotent space} \(X\) is a path-connected space such that \(\pi_1(X)\) is a nilpotent group\footnote{Given a group \(G\), let the \(i\)th-order centres \(\operatorname Z_i(G)\) be the subgroups of \(G\) defined as \(\operatorname Z_0(G)=\{1_G\}\) and \(\operatorname Z_{i+1}(G)=\{x\in G|\forall y\in G\colon xyx^{-1}y^{-1}\in\operatorname  Z_i(G)\}\). A group \(G\) is nilpotent if \(\operatorname Z_i(G)=G\) for sufficiently large \(i\).} and its action on \(\pi_k(X)\) for \(k>1\) is nilpotent\footnote{
   A group action \(\phi\colon G\to\operatorname{Sym}(S)\) is nilpotent if the image \(\phi(G)\le\operatorname{Sym}(S)\) is a nilpotent group.}.
\end{definition}
Consider the category \(\operatorname{Top_{nilp}}\) of nilpotent spaces with finite Betti numbers and continuous maps between them. Now, \(\operatorname{Top_{nilp}}\) can be promoted to an \(\infty\)-category; more precisely, there exists a model-category structure on \(\operatorname{Top_{nilp}}\) in which the weak equivalences are rational homotopy equivalences.

Now, the following theorem tells us that a rational homotopy type is essentially the same as a dgc-algebra in positive degrees.
\begin{theorem}[{{\cite[Theorem~2.35]{felix}}}]
    There exists a Quillen equivalence of model categories between \(\operatorname{dgcAlg}_{>0}\) and
    the opposite category of \(\operatorname{Top_{nilp}}\).
\end{theorem}
In particular, to any rational homotopy type of a space belonging to \(\operatorname{Top_{nilp}}\), one can associate a differential graded-commutative algebra in \(\operatorname{dgcAlg}_{>0}\) uniquely up to quasi-isomorphism, and vice versa.
The construction proceeds by associating to each nilpotent topological space \(X\) with finite Betti numbers a `rational version' of the de~Rham complex.
While differential forms on an ordinary smooth manifold can only be defined over the real numbers,
on a \emph{simplicial set} (such as one obtained by a triangulation of a manifold) one can naturally define differential forms over the rational numbers; and any topological space can be represented by a simplicial set (up to weak homotopy equivalence).

\paragraph{Computation of ranks of homotopy and cohomology groups.}
The dgc-algebra \(A=\Omega^{\mathbb Q}(X)\) corresponding to a topological space \(X\) encodes all information about the rational homotopy type and in particular knows about the ranks of the homotopy and cohomology groups: the cohomology groups can be read off from the \emph{cohomology} of \(A\) while the homotopy groups can be read off from the \emph{Sullivan model} of \(A\).
\begin{theorem}[{{\cite[Theorem~2.50]{felix}}}]
    Give a nilpotent space \(X\) with finite Betti numbers whose corresponding dgc-algebra is \(A\), the \(i\)\textsuperscript{th} Betti number (the rank of the \(i\)\textsuperscript{th} cohomology group) of \(X\) is given by \(\dim\operatorname H^i(A)\), and the rank of the \(i\)\textsuperscript{th} homotopy group of \(X\) is given by the number of degree \(i\) generators of the Sullivan model of \(A\).
\end{theorem}
For instance, consider the \(n\)-sphere \(\mathbb S^n\), whose associated dgca is
\begin{equation}
    A = \mathbb Q[x]/(x^2),\;|x| = n,\;\mathrm dx=0.
\end{equation}
The cohomology of \(A\) is \(A\) itself, so that
\begin{equation}
    \operatorname{rk}\operatorname H^i(\mathbb S^n) = \dim A^i 
    = \begin{cases}
        1 & \text{if \(i\in\{0,n\}\)} \\
        0 & \text{otherwise}.
    \end{cases}
\end{equation}
When \(n\) is odd, \(A\) is its own Sullivan model (since \(x^2=0\) due to degree reasons), but when \(n\) is even, the Sullivan model is
\begin{equation}
    A_\mathrm{Sull} = \mathbb Q[x,y],\;|x|=n,\;|y|=2n-1,\;\mathrm dx=0,\;\mathrm dy=x^2.
\end{equation}
Hence the ranks of the homotopy groups of spheres are
\begin{equation}
    \operatorname{rk}\pi_i(\mathbb S^n)
    =\begin{cases}
        1 & \text{if (\(n\) odd and \(i=n\)) or (\(n\) even and \(i\in\{n,2n-1\}\))} \\
        0 & \text{otherwise}.
    \end{cases}
\end{equation}

Since the Sullivan algebra is quasifree, we can take the corresponding dual \(L_\infty\)-algebra.
\begin{definition}
    Given a nilpotent space \(X\) with finite Betti numbers, the \emph{Whitehead \(L_\infty\)-algebra} \(\mathfrak l(X)\) is the \(L_\infty\)-algebra such that \(\operatorname{CE}(\mathfrak l(X))\) is the Sullivan model for the dgca corresponding to \(X\).
\end{definition}
The underlying graded vector space of \(\mathfrak l(X)\) is therefore
\begin{equation}
    \mathfrak l(X) = \bigoplus_{i=1}^\infty (\pi_i(X)\otimes\mathbb Q)[i-1].
\end{equation}
Of course, if desired, one can tensor with another field to obtain a real \(L_\infty\)-algebra \(\mathfrak l(X)\otimes_{\mathbb Q}\mathbb R\) or a complex \(L_\infty\)-algebra \(\mathfrak l(X)\otimes_{\mathbb Q}\mathbb C\).

\subsection{Whitehead towers}\label{ssec:whitehead-tower}
Suppose that we have a Sullivan model \(A=A_{(0)}\) whose generators are ordered as \(x_1,x_2,\dotsc\). By definition, \(\mathrm dx_1=0\) and defines an element of the nontrivial homotopy group of lowest degree. We can trivialise this by adjoining a new generator \(y\) such that \(\mathrm dy=x_1\); then \(x_1\) and \(y\) form a trivial pair. Let \(\tilde A\) be the Sullivan model of the resulting algebra; then we can apply the previous construction to the first generator of \(\tilde A\), and so on. Iterating, we can construct a chain of Sullivan models
\begin{equation}
    A_{(0)}\to A_{(1)}\to A_{(2)}\to\dotsb
\end{equation}
such that \(A_{(1)}\) is simply connected (i.e.\ has no generators of degree \(\le1\)), \(A_{(2)}\) is two-connected (has no generators of degree \(\le2\)), and so on. Taking the limit, one obtains a trivial Sullivan model corresponding to a contractible space since we have iteratively killed the rational homotopy groups.

\begin{subequations}\label{eq:whitehead-example}
As a simple example, consider the Sullivan dgca
\begin{equation}A = A_{(1)} = \mathbb Q[x_2, x_3]\end{equation}
with \(|x_i| = i\) and \(\mathrm dx_3 = (x_2)^2\) and \(\mathrm dx_2=0\), describing the rational homotopy type of the two-sphere \(\mathbb S^2\). It has a nontrivial rational second homotopy group corresponding to the generator \(x_2\).
Therefore, we can introduce a degree-one generator \(x_1\) with \(\mathrm x_1 = x_2\) to trivialise \(x_2\); afterwards, the new dgca has nontrivial cohomology in degree 3 spanned by \(y_3=x_3 - x_1 x_2\). The Sullivan model of this new dgca is
\begin{equation}
    A_{(2)}=\mathbb Q[y_3]
\end{equation}
with \(|y_3|=3\) and with trivial differential; this is the rational homotopy type of \(\mathbb S^3\). Similarly, trivialising \(y_3\) by means of adjoining a degree-two generator \(y_2\) with \(\mathrm dy_2=y_3\), the next entry in the chain
\begin{equation}
    A_{(3)}=\mathbb Q
\end{equation}
is trivial, modelling the rational homotopy type of a contractible space.
\end{subequations}

This construction has an analogue in ordinary homotopy theory called the \emph{Whitehead tower} \cite[Example~4.20]{hatcher} (in contrast to which the rational-homotopy-theoretic construction may be called the \emph{rational Whitehead tower}), where one iteratively trivialises homotopy groups including any torsion components; an explicit construction may be given in terms of cellular complexes. That is, starting with a pointed homotopy type \(X=X_{(-1)}\), one has the sequence of continuous maps
\begin{equation}
    X_{(-1)}\xleftarrow{\text{trivialise \(\pi_0\)}} X_{(0)}
    \xleftarrow{\text{trivialise \(\pi_1\)}} X_{(1)}
    \xleftarrow{\text{trivialise \(\pi_2\)}} X_{(2)}
    \xleftarrow\dotsb,
\end{equation}
such that \(X_{(0)}\) is connected, \(X_{(1)}\) is simply connected, \(X_{(2)}\) is two-connected (has trivial homotopy groups of degree \(\le2\)), and so on; the colimit of this Whitehead tower is therefore contractible. Note that the first step is simply taking the connected component (containing the basepoint) and the second step is taking the universal covering space; in particular, if \(X_{(-1)}\) is a manifold, then \(X_{(0)}\) and \(X_{(1)}\) remain manifolds of the same dimension. However, this is no longer the case for \(X_{(2)}\) and beyond.

We can now reinterpret the example \eqref{eq:whitehead-example} as the rational version of the Whitehead tower of \(X=\mathbb S^2\), which proceeds as
\begin{multline}
    \mathbb S^2 \xleftarrow{\text{trivialise \(\pi_2(\mathbb S^2)=\mathbb Z\)}}
    X_{(2)} \xleftarrow{\text{trivialise \(\pi_3(X_{(3)})=\mathbb Z\)}}\\
    X_{(3)} \xleftarrow{\text{trivialise \(\pi_4(X_{(3)})=\mathbb Z\)}}
    X_{(4)} \leftarrow\dotsb.
\end{multline}
Here, \(A\) is the Sullivan model of \(X = \mathbb S^2\) while $A_{(2)}$ is the Sullivan model of \(X_{(2)} = \mathbb S^3\) and \(A_{(3)}\) is the Sullivan model of \(X_{(3)}\).
The subsequent homotopy groups after \(\pi_4(X_{(3)})\) are all torsion, so the rational Whitehead tower terminates at \(A_{(3)}\) whereas the non-rational Whitehead tower continues.

Let us now consider the orthogonal group \(\operatorname O(n)\). Its homotopy groups \(\pi_i(\operatorname O(n))\) are, for \(i\le n-2\), given by Bott periodicity with period eight:
\begin{equation}
    \pi_i(\operatorname O(n)) = \begin{cases}
        \mathbb Z_2 &\text{if \(i\equiv 0,1\pmod8\)}\\
        \mathbb Z &\text{if \(i\equiv 3,7\pmod8\)}\\
        0 &\text{otherwise}
    \end{cases}\qquad(i\le n-2).
\end{equation}
For \(n\) sufficiently large, therefore, we obtain the following Whitehead tower \cite{Sati:2009ic}:
\begin{multline}
    \operatorname O(n) \xleftarrow{\text{trivialise \(\pi_0\)}}
    \operatorname{SO}(n) \xleftarrow{\text{trivialise \(\pi_1\)}}
    \operatorname{Spin}(n) \xleftarrow{\text{trivialise \(\pi_3\)}}\\
    \operatorname{String}(n) \xleftarrow{\text{trivialise \(\pi_7\)}}
    \operatorname{Fivebrane}(n) \xleftarrow{\text{trivialise \(\pi_8\)}}
    \operatorname{SFivebrane}(n) \xleftarrow{\text{trivialise \(\pi_9\)}}\\
    \operatorname{SpinFivebrane}(n) \xleftarrow{\text{trivialise \(\pi_{11}\)}}
    \operatorname{Ninebrane}(n)\leftarrow
    \dotsb,
\end{multline}
successively obtaining the special orthogonal group \(\operatorname{SO}(n)\),
the spin group \(\operatorname{SO}(n)\),
the string group \(\operatorname{String}(n)\),
the fivebrane group \(\operatorname{Fivebrane}(n)\) (as named in \cite{Sati:2009ic}),
and what we may call the special fivebrane group \(\operatorname{SFivebrane}(n)\) and the spin fivebrane group \(\operatorname{SpinFivebrane}(n)\)\footnote{These two names are not standard.},
followed by what has been called the ninebrane group \(\operatorname{Ninebrane}(n)\) in \cite{Sati:2009ic}.
The first three are finite-dimensional Lie groups, but starting with the string group, these stop being finite-dimensional Lie groups.

In rational homotopy theory, the non-torsion part of the above Whitehead tower reduces to the following series of \(L_\infty\)-algebras:
\begin{equation}
    \mathfrak o(n) \xrightarrow{\text{trivialise \(\pi_3\otimes\mathbb Q\)}}
    \mathfrak{string}(n) \xrightarrow{\text{trivialise \(\pi_7\otimes\mathbb Q\)}}
    \mathfrak{ninebrane}(n)\to\dotsb,
\end{equation}
where at each stage we add a single \(L_\infty\)-algebraic cocycle; these are all instances of what has been called the `Chern--Simons \(L_\infty\)-algebra' construction in \cite[Def.~23]{Sati:2008eg}.

\section{The Sati--Schreiber charge-quantisation postulate and its refinement}\label{sec:quantisation}
\subsection{Sati--Schreiber flux quantisation}\label{ssec:sati-schreiber-flux-quantisation}
On a globally hyperbolic spacetime \(M\) diffeomorphic to \(\Sigma\times\mathbb R\), a Maxwell-type higher gauge theory is described on a Cauchy surface \(\Sigma\) by its flux densities \(B=(B^i)\) which satisfy a higher Gauss law
\begin{equation}
    \mathrm d_\Sigma B^i = P^i(B)
\end{equation}
for some set of polynomials\footnote{that is, polynomials in graded-commutative variables, where the graded-commutative product is the wedge product of differential forms} \(P^i\).
Sati and Schreiber \cite{Sati:2023mta,Sati:2024klw,Sati:2025vjw} observed that the collection of polynomials \(P^i\) may be encoded by a characteristic \(L_\infty\)-algebra \(\mathfrak a\) over the real numbers, known as the \emph{Gauss-law \(L_\infty\)-algebra}, defined by requiring that its Chevalley--Eilenberg algebra be generated by symbols \(\mathtt b=(\mathtt b^i)\) of the same degrees as the fluxes, with differential
\begin{equation}
    \mathrm{d}_\mathrm{CE}\mathtt b^i = P^i(\mathtt b).
\end{equation}
In this way, flux configurations on \(\Sigma\) are equivalently flat \(\mathfrak a\)-valued differential forms. Thus \(\mathfrak a\) is the algebraic object that encodes the local higher Gauss law obeyed by the flux sector on a Cauchy surface.

\begin{postulate}[note=Sati--Schreiber charge quantisation \cite{Sati:2024klw,Sati:2025vjw}]\label{post:unrefined-charge-quantisation}
A flux-quantisation law for a theory characterised by a Gauss-law \(L_\infty\)-algebra \(\mathfrak{a}\) is a choice of homotopy type \(\mathcal{A}\) together with a quasi-isomorphism \(\mathfrak a \simeq \mathfrak l(\mathcal A)\otimes_{\mathbb Q}\mathbb R\),
where the right-hand side carries the structure of a real \(L_\infty\)-algebra given by the Whitehead brackets.
\end{postulate}

Such a choice of \(\mathcal A\) provides a global topological refinement of the local flux data on \(\Sigma\), which endows solutions of the higher Gauss law with discrete charge sectors.

For example, in ordinary Maxwell theory in four dimensions, the flux densities are the electric and magnetic fields \(B=(\mathbf E, \mathbf B)\) satisfying the Gauss laws
\begin{equation} 
\mathrm{d}_\Sigma\mathbf E=0, \quad  \mathrm{d}_\Sigma\mathbf B=0,
\end{equation}
so that the Gauss-law \(L_\infty\)-algebra is the Abelian differential graded Lie algebra \(\mathfrak a = \mathbb R[1]\oplus\mathbb R[1]\) whose brackets and differential are zero. A natural choice is then \(\mathcal A = \mathrm B^2\mathbb Z\times\mathrm B^2\mathbb Z=\mathbb{CP}^\infty\times\mathbb{CP}^\infty\), which corresponds to the usual Dirac quantisation condition for electric and magnetic charges.

A choice of \(\mathcal{A}\) yields a corresponding notion of quantised charge on the Cauchy surface \(\Sigma\): a discrete charge sector for a theory is given by an element of
\begin{equation}
    \operatorname H^1(\Sigma;\Omega \mathcal A) \cong \pi_{0}\mleft(\operatorname{Map}(\Sigma,\mathcal A)\mright),
\end{equation}
which is the set of homotopy classes of maps \(\Sigma\rightarrow \mathcal A\). (Here, \(\Omega\) denotes the based loop space.) Now, since \(\mathfrak{a}\simeq \mathfrak l(\mathcal A)\otimes_{\mathbb Q}\mathbb R\), there is a non-Abelian character map
\begin{equation}
    \operatorname{ch}^{\mathcal A}_\Sigma\colon \operatorname H^1(\Sigma;\Omega\mathcal A)\longrightarrow \operatorname H^1_{\mathrm{dR}}(\Sigma;\mathfrak a)
\end{equation}
into de~Rham cohomology with coefficients in \(\mathfrak a\), which sends a charge class to the corresponding total flux class. Thus, a solution of the higher Gauss law is said to satisfy the flux quantisation law determined by \(\mathcal A\) precisely when its total flux class \([B]\in\operatorname H^1_{\mathrm{dR}}(\Sigma;\mathfrak a)\) admits a lift to a class in \(\operatorname H^1(\Sigma;\Omega\mathcal A)\). In this way, the choice of \(\mathcal A\) refines the local differential-form data of the flux density by a global discrete charge.  

From our perspective, the proposal of Sati and Schreiber  captures a crucial structural point: the local flux data encoded by a Gauss-law \(L_\infty\)-algebra \(\mathfrak a\) generally admits a further global topological refinement, encoded by a homotopy type \(\mathcal A\), and it is this additional datum that governs charge quantisation \cite{Sati:2024klw,Sati:2025vjw}. In this sense, the homotopy type \(\mathcal A\) should be regarded as part of the global definition of the physical model, rather than as something visible from the local differential equations alone:
    \begin{quote}
    Therefore, the choice of \(\mathcal{A}\) is part of the specification of the \emph{physical model}. If the physics to be described by that model is thought to be fixed, then the choice of \(\mathcal{A}\) is a \emph{hypothesis} about the correct mathematical description of that physics, each such choice entailing \emph{predictions} about the global behaviour of the given physical fields. [\cite{Sati:2025vjw}, emphasis in original.]
    \end{quote}

However, the above formulation can be developed further in some interesting directions.
In fact, as stated, the framework applies to flux sectors admitting a higher Maxwell-type description in terms of a Gauss-law \(L_\infty\)-algebra on a Cauchy surface. Extending this picture to genuinely non-Abelian gauge theories is an important problem, and part of the motivation for the refinement developed in the next subsection.
In the non-Abelian case, it is not immediately clear what should play the role of the \(L_\infty\)-algebra \(\mathfrak a\), since neither the field strengths nor the Gauss law are, in general, gauge-invariant. A first step is therefore to construct an \(L_\infty\)-algebra of gauge-invariant fluxes satisfying equations that generalise the Gauss law of Maxwell-type theories.

Once such an algebra has been identified, the choice of \(\mathcal A\) remains constrained by the local physics encoded by \(\mathfrak a\) (though in general not completely determined by it) and hence by the kinematics and dynamics of the theory. In many examples, one is naturally led to choose \(\mathcal A\) so that it captures the homotopy type of the \emph{space of kinematically allowed fields} in the following sense. Let \(M\) be a globally hyperbolic Lorentzian manifold with Cauchy surface \(\Sigma \subset M\), and let \(\mathcal C(\Sigma)\) denote the space of admissible initial data on \(\Sigma\), namely the values of the fields together with their conjugate momenta, subject to the Gauss-law and other constraints. At the level of homotopy type, \(\mathcal C(\Sigma)\) may be viewed as the phase space of the theory. In the examples considered here, one is then led to seek a homotopy equivalence
\begin{equation}
    \mathcal C(\Sigma) \simeq \mathrm{Map}(\Sigma,\mathcal A),
\end{equation}
so that \(\mathcal A\) may be regarded as as encoding the topological sectors of admissible initial data.
In Yang--Mills theory, for example, \(\mathcal A\) must at least contain \(\mathrm{B}H\), the classifying space of principal \(H\)-bundles, where \(H\) is the gauge group. This reflects the fact that field configurations consisting of a connection \(A\) and its canonical conjugate momentum \(\piit\), subject to the Gauss law, and built on topologically inequivalent principal \(H\)-bundles over \(\Sigma\), cannot be continuously deformed into one another. In many theories, however (including Yang--Mills theory, as discussed in \cref{sec:yang-mills}) this does not exhaust all choices of \(\mathcal A\) compatible with the local physics. Indeed, taking only \(\mathcal A = \mathrm{B}H\) would capture topologically magnetic charges, but would leave no room for electric-type charges to be incorporated into the charge-quantisation law.

Furthermore, the discussion in \cite{Sati:2024klw,Sati:2025vjw} is primarily restricted to the gauge sector of the theory: \(\mathcal A\) describes gauge-invariant field strengths and not (for instance) gauge-invariant combinations of matter fields. From a more general point of view, however, there is no need to discriminate between gauge fields and matter fields.
In particular, currents that couple magnetically to gauge fields naturally appear in Bianchi identities: for instance, in Maxwell theory, one has \(\mathrm dF=J\), where \(F\) is the two-form field strength and \(J\) a three-form magnetic current. 

Finally, the formulation in \cref{post:unrefined-charge-quantisation} is given in terms of \(L_\infty\)-algebras defined over the field \(\mathbb R\) of real numbers.
However, in rational homotopy theory, the Whitehead brackets are naturally defined over the \emph{rational} homotopy groups \(\pi_\bullet(\mathcal A)\otimes\mathbb Q\), and the rational homotopy type is a finer classification than the real homotopy type.
Similarly, in all natural examples occurring in physics, the algebra \(\mathfrak a\) (and, more generally, \(\mathfrak h\) and \(\mathfrak w\)) can always be defined over the rational numbers. For instance, the structure constants of simple compact Lie algebras are always rational numbers.

In order to address these aspects, we first introduce the basic framework of adjusted higher gauge theories and develop a language for their fluxes.

\subsection{Adjusted higher gauge theories and their fluxes}
Consider a gauge theory living on a spacetime \(M\) with gauge symmetry \(\mathfrak h\), where \(\mathfrak h\) is a Lie algebra with an invariant inner product or, more generally, an \(L_\infty\)-algebra.
On a local patch \(U\subset M\), which we assume to be contractible, there are two natural differential graded-commutative algebras: the Chevalley--Eilenberg algebra \(\operatorname{CE}(\mathfrak h)\) and the de~Rham algebra of differential forms \(\Omega^\bullet(U)=\mathcal C^\infty(\mathrm T[1]U)\), which can be regarded as the Chevalley--Eilenberg algebra of the tangent Lie algebroid \(\mathrm TU\) of \(U\). It is then natural to consider the set of dgca morphisms
\begin{equation}
    \operatorname{CE}(\mathfrak h)\to\Omega(U).
\end{equation}
Such a dgca morphism describes flat \(\mathfrak h\)-valued connections. For instance, suppose that \(\mathfrak h=\mathfrak{su}(2)\), spanned by \(\mathtt t_1,\mathtt t_2,\mathtt t_3\) (so that \(\operatorname{CE}(\mathfrak g)\) is generated by the dual basis \(\mathtt t^1,\mathtt t^2,\mathtt t^3\)). Then the Chevalley--Eilenberg algebra \(\operatorname{CE}(\mathfrak g)\) is
\begin{align}
    \mathrm d \mathtt t^1 &= - \mathtt t^2\wedge\mathtt t^3,&
    \mathrm d \mathtt t^2 &= - \mathtt t^3\wedge\mathtt t^1,&
    \mathrm d \mathtt t^3 &= - \mathtt t^1\wedge\mathtt t^2,
\end{align}
so that a dgca morphism \(A\colon\operatorname{CE}(\mathfrak{su}(2))\to\Omega(U)\) consists of the data \(A(\mathtt t^i)=A^i\) with
\begin{align}
    \mathrm dA^1 &= -A^2\wedge A^3,&
    \mathrm dA^2 &= -A^3\wedge A^1,&
    \mathrm dA^3 &= -A^1\wedge A^2,
\end{align}
i.e.\ defining \(A=A^i\mathtt t_i\in\Omega(U;\mathfrak{su}(2))\) and \(F=\mathrm dA+\frac12[A,A]\), then
\begin{equation}
    F = 0.
\end{equation}
The culprit is the fact that \(\operatorname{CE}(\mathfrak h)\) does not include generators for the field strengths.
In order to describe general (i.e.\ not necessarily flat) connections, one must enlarge \(\mathfrak h\) to a `doubled' \(L_\infty\)-algebra, on the underlying graded vector space \(\mathfrak h\oplus\mathfrak h[1]\).
One canonical way of doing so is called the \emph{inner-derivation \(L_\infty\)-algebra} \(\mathfrak{inn}(\mathfrak h)\) \cite{Sati:2008eg}; its Chevalley--Eilenberg algebra  \(\operatorname{CE}(\mathfrak{inn}(\mathfrak h))\) is classically known as the Weil algebra \cite{SHC_1949-1950__2__A18_0} in the case where \(\mathfrak h\) is a Lie algebra. For instance, \(\operatorname{CE}(\mathfrak{inn}(\mathfrak{su}(2)))\) has differentials
\begin{equation}
\begin{aligned}
    \mathrm d\mathtt t^1 &= - \mathtt t^2\mathtt t^3+\mathtt f^1,&
    \mathrm d\mathtt t^2 &= - \mathtt t^3\mathtt t^1+\mathtt f^2,&
    \mathrm d\mathtt t^3 &= - \mathtt t^1\mathtt t^2+\mathtt f^3,\\
    \mathtt d\mathtt f^1 &= \mathtt f^2 \mathtt t^3 - \mathtt t^2\mathtt f^3,&
    \mathtt d\mathtt f^2 &= \mathtt f^3 \mathtt t^1 - \mathtt t^3\mathtt f^1,&
    \mathtt d\mathtt f^3 &= \mathtt f^1 \mathtt t^2 - \mathtt t^1\mathtt f^2,&
\end{aligned}
\end{equation}
where we have adjoined generators \(\mathtt f^i\) of degree one corresponding to \(\mathfrak h[1]^*\), so that for a dgca morphism \(A\colon \operatorname{CE}(\mathfrak{inn}(\mathfrak{su}(2)))\to\Omega(U)\), if one defines
\(A=A(\mathtt t^i)\mathtt t_i\in\Omega^1(U;\mathfrak{su}(2))\) and \(F=A(\mathtt f^i)\mathtt t_i\in\Omega^2(U;\mathfrak{su}(2))\), then one has
\begin{align}
    \mathrm dA&=-\frac12[A\wedge A]+F,&
    \mathrm dF&=-[A,F]
\end{align}
as expected, and \(F\) is the Yang--Mills field strength. The inner-derivation algebra \(\mathfrak{inn}(\mathfrak h)\) comes with a canonical inclusion \(\mathfrak h\hookrightarrow\mathfrak{inn}(\mathfrak h)\) and therefore a short exact sequence
\begin{equation}\label{eq:ses-BH}
    \mathfrak h\hookrightarrow\mathfrak{inn}(\mathfrak h)\twoheadrightarrow\mathfrak{inv}(\mathfrak h),
\end{equation}
where \(\mathfrak{inv}(\mathfrak h)\) is the \(L_\infty\)-algebra of invariant polynomials; in the case where \(\mathfrak h\) is a semisimple Lie algebra, this is nothing other than the Abelian \(L_\infty\)-algebra on the graded vector space of invariant polynomials of the Lie algebra in the classical sense. A dgca morphism
\begin{equation}
    \operatorname{CE}(\mathfrak{inv}(\mathfrak h))\to\Omega(U)
\end{equation}
that factors as
\begin{equation}
    \operatorname{CE}(\mathfrak{inv}(\mathfrak h))\to\operatorname{CE}(\mathfrak{inn}(\mathfrak h))\to\Omega(U),
\end{equation}
then corresponds to gauge-invariant local observables of the gauge potentials. For instance, if \(\mathfrak h=\mathfrak{su}(N)\), then \(\mathfrak{inv}(\mathfrak{su}(N))=\mathbb R[3]\oplus\mathbb R[5]\oplus\dotsb\oplus\mathbb R[2N-1]\), so that the morphism \(\operatorname{CE}(\mathfrak{inv}(\mathfrak h))\to\Omega(U)\) corresponds to the collection of differential forms \(\operatorname{tr}(F^2),\dotsc,\operatorname{tr}(F^N)\).

Furthermore, if \(H\) is a Lie group integrating \(\mathfrak h\), in favourable cases (e.g.\ if \(\mathfrak h\) is a simple Lie algebra and \(H\) is compact and simply connected), \eqref{eq:ses-BH} is the real homotopy type of the short exact sequence
\begin{equation}
    H\hookrightarrow\mathrm EH\twoheadrightarrow\mathrm BH
\end{equation}
of the classifying space of \(H\).

\paragraph{Adjusted fluxes.}
This canonical construction, however, sometimes fails to be physically suitable. In some cases this is due to the closure of gauge symmetries and the requirement for non-trivial non-Abelian kinematics \cite{Gastel:2018joi,Saemann:2019leg,Samann:2019eei}.
In other cases, the presence of magnetic currents modifies the Bianchi identities.
These magnetic currents can be matter fields or sometimes other gauge fields in theories such as ten-dimensional supergravities.
In these cases, the unmodified Bianchi identities produced by the canonical \(\mathfrak{inn}(\mathfrak h)\) fail to be suitable for describing the physics.
Instead, one takes an \emph{adjusted inner-derivation Lie algebra} \cite{Sati:2008eg,Samann:2019eei,Sati:2009ic,Schmidt:2019pks,Kim:2019owc,Borsten:2021ljb,Tellez-Dominguez:2023wwr,Fischer:2024vak,Gagliardo:2025oio}  (reviewed in \cite{Borsten:2024gox}) \(\mathfrak w\).
If we do not include other currents, then \(\mathfrak w\) will have the same underlying graded vector space \(\mathfrak h\oplus\mathfrak h[1]\) as \(\mathfrak{inn}(\mathfrak h)\) but has different brackets;
if we include additional currents, then the underlying graded vector space of \(\mathfrak w\) will be larger.\footnote{
    This is more general than in previously cited works, where \(\mathfrak w\) does not include additional currents.
}
Then we can construct a short exact sequence
\begin{equation}
    \mathfrak h\hookrightarrow\mathfrak w\twoheadrightarrow\mathfrak a,
\end{equation}
where the quotient \(\mathfrak w/\mathfrak h\eqqcolon\mathfrak a\) may be termed the \emph{adjusted invariant polynomials} of \(\mathfrak h\). The data of \(\mathfrak w\) and \(\mathfrak a\) are \emph{not} determined by \(\mathfrak h\) alone but requires additional data, the so-called \emph{adjustment}.

By construction, then, the \(L_\infty\)-algebra \(\mathfrak a\) consists only of gauge-invariant flux observables \(\mathtt f_i\) whose Chevalley--Eilenberg algebra \(\operatorname{CE}(\mathfrak a)\) encodes Bianchi identities of the form
\begin{equation}
    \mathrm d\mathtt f^i = -f^i_j\mathtt f^j - \frac12f^i_{jk}\mathtt f^j\mathtt f^k-\frac16f^i_{jkl}\mathtt f^j\mathtt f^k\mathtt f^l-\dotsb,
\end{equation}
where `Bianchi identity' is taken in a generalised sense as an equation relating the exterior derivative of a gauge-invariant field (e.g.\ Abelian field strengths, Abelian currents, invariant polynomials of covariant Yang--Mills field strengths and covariant currents) to other gauge-invariant fields.
This provides a natural generalisation of the Gauss-law \(L_\infty\)-algebra of Maxwell-type theories, and therefore \(\mathfrak a\) is the natural candidate for the algebraic object encoding the local flux data of a gauge theory with gauge algebra \(\mathfrak h\).
In particular, in the Maxwell-type case, the Bianchi identities reproduce the Gauss laws (see subsection \ref{sec:type-II-m-theory} for the adjusted gauge theory formulation of the $C$-field in M-theory).
In the case of Yang--Mills theory without matter (with gauge group \(\operatorname{SU}(N)\) for definiteness), \(\mathfrak a\) is generated by the invariant polynomials \(\operatorname{tr}(F^2),\dotsc,\operatorname{tr}(F^N)\) with zero differential, so that the Bianchi identities are simply \(\mathrm d\operatorname{tr}(F^k)=0\).
In the non-Abelian case, these equations do not imply the Gauss laws; nevertheless, we argue that they still encode local flux data.

Given these definitions, we postulate the following small refinement of the original charge-quantisation postulate of \cite{Sati:2025vjw}.
\begin{postulate}[manual-num={1\('\)},note=Refined Sati--Schreiber charge quantisation]\label{post:charge-quantisation}
In a gauge theory with gauge algebra \(\mathfrak h\) and adjusted inner-derivation algebra \(\mathfrak w\hookleftarrow \mathfrak{h}\), we define the \(L_\infty\)-algebra of fluxes to be \(\mathfrak a \coloneqq \mathfrak w/\mathfrak h\). 
Furthermore, \(\mathfrak h\) and \(\mathfrak w\) (and thus \(\mathfrak a\)) can be defined over \(\mathbb Q\) rather than \(\mathbb R\).
Then, a charge-quantisation law for such a theory is a choice of homotopy type \(\mathcal A\) together with a quasi-isomorphism 
\begin{equation}
    \mathfrak a \simeq \mathfrak l(\mathcal A)
\end{equation}
where \(\mathfrak l(\mathcal A) = \pi_\bullet(\mathcal A)\otimes\mathbb Q\) is the graded \(\mathbb Q\)-vector space of rational homotopy groups of \(\mathcal A\) equipped with the Whitehead brackets.
On a Cauchy surface \(\Sigma\subset M\) of spacetime, the homotopy type of the phase space of the theory is \(\mathrm{Map}(\Sigma,\mathcal A)\). 
\end{postulate}

\paragraph{The Hodge pairing.}
In a non-topological theory, the algebra \(\mathfrak a\) often comes with a pairing
\begin{equation}\label{eq:hodge-star-fraka}
    \langle-,-\rangle\colon \mathfrak a\times\mathfrak a\to\mathbb R[d-2]
\end{equation}
that derives from the Hodge star. For instance, in \((p-1)\)-form electrodynamics with multiple \(p\)-form field strength \(F\) valued in a vector space \(V\) and its \((d-p)\)-form dual field strength \(G=*F\) (also valued in vector space), the algebra \(\mathfrak a\) is
\begin{equation}
    \mathfrak a = V[p-1]\oplus V[d-p-1]
\end{equation}
with an evident pairing
\begin{equation}
    \mathfrak a\times\mathfrak a\to\mathbb R[d-2]
\end{equation}
corresponding to
\begin{equation}
    (F^i,G_i)\mapsto \int_M F^i\wedge G^j\delta_{ij},
\end{equation}
where \(\delta_{ij}\) are the coefficients of an inner product on \(V\) that is used in the kinetic term \(\int -F^i\wedge\star F^j\delta_{ij}\).
Phrased in terms of \(\mathcal A\), \eqref{eq:hodge-star-fraka} becomes
\begin{equation}
    (\pi_i(\mathcal A)\otimes\mathbb Q)\times(\pi_{d-i}(\mathcal A)\otimes\mathbb Q) \to \mathbb Q.
\end{equation}
It is a natural question to ask if this lifts to the torsion components of \(\pi_\bullet(\mathcal A)\), i.e. whether a putative map
\begin{equation}
    \pi_i(\mathcal A)\times\pi_{d-i}(\mathcal A)\xrightarrow? \mathbb Z
\end{equation}
should exist. However, this does not appear to be the case. For instance, for \(\mathcal N=4\) Yang--Mills theory in \(d=4\), we have one-form electric and magnetic symmetries valued in \(\operatorname Z(G)\) and \(\pi_1(G)\) for gauge group \(G\), and both of these correspond to components of \(\pi_2(\mathcal A)\). However, there is no natural pairing between \(\operatorname Z(G)\) and \(\pi_1(G)\).

\paragraph{Remark on rationality.}
Let us remark on the reason for formulating \cref{post:charge-quantisation} over \(\mathbb Q\). Locally, the fluxes are real differential forms, and thus determine a real \(L_\infty\)-algebra structure of gauge-invariant fluxes. The point is that charge quantisation requires more than this real local model: it requires a global homotopy type \(\mathcal A\) whose rational Whitehead algebra $\mathfrak{l}(\mathcal{A})$ recovers the local flux algebra after extension of scalars. Requiring the existence of such a rational form is therefore a first step towards an integral, charge-quantised refinement. 

In this sense, the rationality requirement already plays a swampland-like role. It rules out putative real flux \(L_\infty\)-algebras which can be written locally but cannot arise by extension of scalars from rational homotopy data. Such theories may make sense as systems of formal real differential equations, but they fail the charge-quantisation criterion proposed here.

\paragraph{Remark on differential refinement.}
Let us stress a limitation of our formulation. As explained in subsection \ref{ssec:sati-schreiber-flux-quantisation}, in the Maxwell-type case, the choice of \(\mathcal A\) is accompanied by a character map
\(\operatorname{ch}^{\mathcal A}_\Sigma: \operatorname H^1(\Sigma;\Omega\mathcal A) \rightarrow \operatorname H^1_{\mathrm{dR}}(\Sigma;\mathfrak a)\),
so that flux quantisation of a gauge potential may be expressed as the condition that its de~Rham flux class admits a lift to a topological charge class. In the adjusted non-Abelian setting considered here, we do not construct the differential-cohomological refinement corresponding to our proposed charge quantisation. In fact, once one passes to the adjusted quotient \(\mathfrak a=\mathfrak w/\mathfrak h\), understood as a local rational model for gauge-invariant flux densities, it is not clear how to generalise the usual homotopy-pullback construction \cite{Fiorenza:2020dcp} of differential cohomology from the character map  for the general adjusted non-Abelian case. Thus, our construction should be viewed as identifying the topological classifying space of charges associated with the adjusted flux algebra. The construction of the corresponding differential theory is beyond the scope of this work.

\subsection{Branes from charge quantisation}\label{ssec:brane}
Given that \(\mathcal A\) plays a crucial role in the global formulation of the theory, it is natural to expect that homotopy invariants of \(\mathcal A\) encode corresponding global features of the physics.
From the perspective of rational homotopy theory, as reviewed in \cref{ssec:rational-homotopy-theory}, the basic invariants are the homotopy groups and the cohomology groups: the torsion-free part of \(\pi_\bullet(\mathcal A)\) is reflected in the generators of the Sullivan model, while the torsion-free part of \(\operatorname H^\bullet(\mathcal A;\mathbb Q)\) is reflected in its minimal model.

The idea that the homotopy groups classify the branes is explicit in work by Sati--Schreiber \cite{Sati:2025vjw,Fiorenza:2019ckz}.
Topologically, the charge of a \(p\)-brane in a \(d\)-dimensional theory is detected by the behaviour of the fields on a sphere linking its worldvolume. Indeed, a \(p\)-brane has worldvolume dimension \(p+1\), so its transverse space has dimension \(d-p-1\), and a small sphere surrounding the brane is therefore \(\mathbb S^{d-p-2}\). If the theory is charge-quantised by a homotopy type \(\mathcal A\), so that a field configuration determines a map into \(\mathcal A\), then restricting the field to the linking sphere gives a map \(\mathbb S^{d-p-2}\rightarrow \mathcal A\).
After fixing the vacuum value at infinity, this may be regarded as a based map, and its homotopy class therefore defines an element of
\begin{equation}
\pi_{d-p-2}(\mathcal A) \cong \pi_0\operatorname{Map}_\ast(\mathbb S^{d-p-2},\mathcal A)
\end{equation}
that is, a connected component of the mapping space from the linking sphere into \(\mathcal A\).
If this class is trivial, the map \(\mathbb S^{d-p-2}\rightarrow \mathcal A\) extends over the ball bounded by the sphere, and the configuration can be continuously deformed to the vacuum in a neighbourhood of the putative brane; if it is non-trivial, this extension is obstructed, and the defect carries a stable topological charge. In this way, the possible \(p\)-brane charges are classified by \(\pi_{d-p-2}(\mathcal A)\) and we have the following claim. 

\begin{postulate}[note=Defect brane charges from charge quantisation]\label{post:branes}
In a \(d\)-dimensional theory, on a space \(\Sigma\), the set of defect conserved charges is given by homotopy classes of maps \(\Sigma\to\mathcal A\). In particular, for every \(p\), the set of possible \(p\)-brane charges is given by \(\pi_{d-p-2}(\mathcal A)\). In particular, if \(\pi_{d-p-2}(\mathcal A)\) has torsion, this corresponds to fractional branes.
\end{postulate}

\paragraph{Defect charges.}
By the term `defect charges', we mean possible charges of defect branes (line operators, surface operators, etc.) that cannot be screened by the \emph{dynamical charges}, which are the charges carried by the dynamical matter or gauge fields present in the theory. For instance, in Maxwell theory without matter, electric charges of defect particles (line operators) are valued in \(\mathbb Z\), but in Maxwell theory with matter \(\phi\) carrying charge \(k\) (so that the theory contains dynamical charges valued in \(k\mathbb Z\)), if \(\phi\) takes a nonzero vacuum expectation value, then charges of defect particles that are multiples of \(k\) can be screened out, so that defect charges are valued in \(\mathbb Z_k\) rather than \(\mathbb Z\). (This example is further discussed in \cref{ssec:maxwell-with-source}.)
    
    The proviso of restricting to defect charges differs from the description given in \cite{Sati:2025vjw,Fiorenza:2019ckz}, which does not contain this restriction.
    This discrepancy stems from whether \(\mathcal A\) describes just the gauge sector of the theory (excluding charged matter), which is the case implicitly assumed by \cite{Sati:2025vjw,Fiorenza:2019ckz}, or whether \(\mathcal A\) describes the entire theory (including charged matter), which is the perspective taken in this paper.

\paragraph{Torsion and non-commutativity.}
For \(k>1\), the torsion-free part of \(\pi_k(\mathcal A)\) is \(\mathbb Z^n\), this means that there are \(n\) distinct types of \((d-k-2)\)-branes. As for the torsion part of \(\pi_k\), a generator of order \(m\) corresponds to \((d-k-2)\)-branes that are `fractional' in the sense that a stack of \(m\) such branes can annihilate; in particular, a configuration with one such brane can be continuously deformed into one with a stack of \(m-1\) anti-branes.

For \(k=1\), the fundamental group \(\pi_1(\mathcal A)\) classifies branes of codimension \(2\), such as seven-branes in string theory. The non-Abelianness of \(\pi_1(X)\) corresponds to the fact that such branes need not be mutually local \cite{Kim:2022opr}. Similarly, \(\pi_0(X)\) classifies domain walls, whose charges need not form a group at all \cite{Kim:2022opr}.

\subsection{Higher symmetries from charge quantisation}
We next explain how the same classifying space \(\mathcal A\) that governs charge quantisation also gives rise to higher-form symmetry operators. The basic idea is that, if field configurations are topologically classified by maps into \(\mathcal A\), then cohomology classes on \(\mathcal A\) can be pulled back to spacetime and evaluated on closed submanifolds, thereby producing topological phases. In this way, the homotopy-theoretic data of \(\mathcal A\) controls not only brane charges, but also invertible Abelian higher-form symmetries, through suitable \(\operatorname U(1)\)-valued cohomology classes. While the interpretation of homotopy groups in terms of brane charges is already explicit in the work of Sati--Schreiber \cite{Fiorenza:2019ckz,Sati:2025vjw}, the corresponding relation between \(\operatorname U(1)\)-valued cohomology and invertible higher-form symmetries appears to be novel, or at least not developed explicitly in the existing literature; although \cite{Sati:2025vjw} mentions generalised symmetries, it does not pursue this point in detail.

Suppose that we are given a theory where (the homotopy type of) the space of solutions is classified by the homotopy type \(\mathcal A\).
Let \([\alpha]\in\operatorname H^k(\mathcal A;\operatorname U(1))\) be a cohomology class. Let \(\iota_\Sigma\colon\Sigma\hookrightarrow M\) be a closed \(k\)-dimensional submanifold of spacetime \(M\).
Given a field configuration \(\phi\) on spacetime, this defines (by assumption) a homotopy class \([\phi]\colon M\to\mathcal A\). Then we can define the operator \(\mathcal O_\alpha(\Sigma)\) that, when inserted into the path integral, takes the value
\begin{equation}\label{eq:path-integral-operator}
    \int\mathrm D\phi\,\mathcal O_\alpha(\Sigma)\exp(\mathrm iS[\phi])
    \coloneqq
    \int\mathrm D\phi\,\left(\int_\Sigma \phi^*\alpha\right)\exp(\mathrm iS[\phi]),
\end{equation}
where the notation \(\int_\Sigma \phi^*\alpha\in\operatorname U(1)\) denotes the tautological pairing of the integral homology class \([\Sigma]\in\operatorname H_k(M)\) with the pullback cohomology class \(\phi^*[\alpha]\in\operatorname H^k(M;\operatorname U(1))\).
This operator is topological by construction; this therefore defines a \((d-k-1)\)-form symmetry valued in the group \(\operatorname H^k(\mathcal A;\operatorname U(1))\).
Since \(\operatorname U(1)\) is a divisible Abelian group (i.e.\ injective module over \(\mathbb Z\)), its Ext groups \(\operatorname{Ext}^\bullet_{\mathbb Z}(-,\operatorname U(1))\) vanish in positive degree, and the universal coefficient theorem for cohomology implies that \(\operatorname H^k(\mathcal A;\operatorname U(1))=\hom(\operatorname H_k(\mathcal A),\operatorname U(1))\) is the Pontryagin dual group of the homology \(\operatorname H_k(\mathcal A)\) with integer coefficients.

Therefore, we have the following result.
\begin{postulate}[note=Higher-form symmetries from charge quantisation]\label{post:higher-form-symmetry}
    For every \(k\), there exist invertible \(k\)-form symmetries valued in
    \begin{equation}\operatorname H^{d-k-1}\mleft(\mathcal A;\operatorname U(1)\mright) \cong \hom\mleft(\operatorname H_{d-k-1}(\mathcal A;\mathbb Z),\operatorname U(1)\mright),\end{equation}
    that is, the Pontryagin dual group of \(\operatorname H_{d-k-1}(\mathcal A;\mathbb Z)\).
\end{postulate}
One can ask whether \emph{every} higher-form symmetry is induced by some element of \(\operatorname H^\bullet(\mathcal A;\operatorname U(1))\). However, note that (the Pontryagin dual groups of) \(\operatorname H^\bullet(\mathcal A;\operatorname U(1))\) can only yield \emph{invertible}, \emph{Abelian} higher-form symmetry groups.

Note that, in the special case when \(\mathcal A\) carries the structure of an H-space,\footnote{An H-space \cite[§3.C]{hatcher} is a `monoid up to homotopy', i.e.\  a pointed topological space \((\mathcal A,\bullet)\) with a continuous basepoint-preserving map \(\mathcal A\times\mathcal A\to\mathcal A\) that is associative up to homotopy and preserves \(\bullet\) as an identity up to homotopy. In particular, a topological monoid is an H-space; the space of unit octonions is an example of an H-space that is not a monoid.}
then \(\operatorname H_\bullet(\mathcal A)\) carries the structure of a graded-commutative ring (the Pontryagin ring).
Thus, for instance, when \(\mathcal A=\mathrm B^kG\) where \(G\) is Abelian, there exists a product structure on the set of charges for invertible higher-form symmetries.

A basic class of examples can be obtained by taking \(\mathcal A = \mathrm BG\), the classifying space of \(G\)-bundles. Since maps \(X\to \mathrm BG\) classify principal \(G\)-bundles on \(X\), the corresponding \(\operatorname U(1)\)-valued cohomology classes define topological phases of \(G\)-bundles. Concretely, a class \([\alpha]\in \operatorname H^k(\mathrm BG; \operatorname U(1))\) assigns to every principal \(G\)-bundle \(P\to \Sigma^k\) on a closed oriented \(k\)-manifold \(\Sigma\) a phase in \(\operatorname U(1)\), obtained by evaluating \([\alpha]\) on the homotopy class \(\Sigma\to \mathrm BG\) classifying \(P\). Thus \(\operatorname H^\bullet(\mathrm BG; \operatorname U(1))\) may be interpreted as the group of \(\operatorname U(1)\)-valued characteristic phases of principal \(G\)-bundles. 

\paragraph{One-form symmetries of Maxwell theory.}
In particular, as we have seen, for four-dimensional Maxwell theory one is naturally led to take \(\mathcal A=\mathrm B^2\mathbb Z\times\mathrm B^2\mathbb Z = \mathrm{BU}(1)\times \mathrm{BU}(1)\), so as to treat the magnetic and electric sectors on the same footing. Since \(\mathrm{BU}(1)\simeq K(\mathbb Z,2)\), one has
\begin{equation}
\operatorname H^2(\mathcal A; \operatorname U(1))\cong \operatorname U(1)\times \operatorname U(1) \cong (\mathbb R/\mathbb Z)^2,
\end{equation}
and a class \((\lambda_\mathrm m,\lambda_\mathrm e)\in (\mathbb R/\mathbb Z)^2\) thus assigns to a pair of principal \(U(1)\)-bundles \((P_\mathrm m,P_\mathrm e)\) on a closed oriented surface \(\Sigma\) the phase
\begin{equation}
\mathcal{O}_{\lambda_\mathrm m,\lambda_\mathrm e}(\Sigma) = \exp\mleft(2\pi\mathrm i\lambda_\mathrm m\int_\Sigma c_1(P_\mathrm m)+2\pi\mathrm i\lambda_\mathrm e\int_\Sigma c_1(P_\mathrm e)\mright),
\end{equation}
where \(c_1(P_\mathrm m)\) and \(c_1(P_\mathrm e)\) are the first Chern classes of the magnetic and electric bundles, respectively. Since the first Chern class of a principal \(U(1)\)-bundle \(P\) is given by the cohomology class of its field strength \(F\) divided by \(2\pi\), we have
\begin{equation}
c_1(P_\mathrm m)=\Big[\frac{F}{2\pi}\Big],\qquad
c_1(P_\mathrm e)=\Big[\frac{\star F}{2\pi}\Big].
\end{equation}
These two factors recover the familiar magnetic and electric one-form symmetries of Maxwell theory: 
\begin{equation}
J_\mathrm m=\frac{F}{2\pi},\qquad \mathrm dJ_\mathrm m=0,
\end{equation}
\begin{equation}
J_\mathrm e=\frac{\star F}{2\pi},\qquad \mathrm dJ_\mathrm e=0.
\end{equation}
The first is the magnetic one-form symmetry, coming from the Bianchi identity \(\mathrm{d}F=0\). The second is the electric one-form symmetry, coming from the sourceless Maxwell equation \(\mathrm{d}{\star}F=0\).

\section{Swampland-type implications of charge quantisation}\label{sec:swampland}
The postulates above imply non-trivial statements about consistency of field theories, which are reminiscent of swampland conjectures about consistent effective field theories that admit ultraviolet completions including gravity. (Of course, these are not literally swampland conjectures in that they do not specifically pertain to effective field theories of quantum gravity.)

\subsection{Gauge symmetries and 't~Hooft-anomaly-free global symmetries must be compact}\label{ssec:swampland-gauge-group}
The charge-quantisation postulate has the consequence of forbidding noncompact gauge symmetries and, therefore, also gaugeable noncompact global symmetries; we give two related arguments.

\paragraph{Discrete non-torsion subgroups are forbidden.}
We first argue that a noncompact gaugeable global symmetry is forbidden because one could gauge a noncompact discrete subgroup of it violating \cref{post:charge-quantisation}.

For simplicity, consider a theory whose flux \(L_\infty\)-algebra is \(\mathfrak a=0\), that is, with no gauge-covariant field strengths. The charge-quantisation postulate requires that the admissible classifying space \(\mathcal{A}\) satisfy \(\mathfrak l(\mathcal A)=0\)
or equivalently, that all the homotopy groups are torsion groups, so that \(\pi_\bullet(\mathcal A)\otimes\mathbb Q=0\). 

Now, let \(G\) be a connected Lie group acting as a 't~Hooft-anomaly-free global symmetry of a theory with \(\mathfrak a=0\) as above. Assume moreover that if a closed subgroup \(\Gamma\subset G\) is given, then \(\Gamma\) may also be gauged. Then \(G\) has to be compact.

To see this, assume for contradiction that \(G\) is connected and noncompact. A standard fact from Lie theory is that every connected noncompact Lie group contains a closed subgroup isomorphic to \(\mathbb R\), and hence also a closed discrete subgroup \(\Gamma\cong \mathbb Z\).
Indeed, if the solvable radical of \(G\) is noncompact, then it contains a closed copy of \(\mathbb R\); if the solvable radical is compact, then the semisimple Levi factor is noncompact, and therefore contains a closed split torus \(\mathbb R^r\) for some \(r\ge 1\). In either case, \(G\) contains a closed copy of \(\mathbb R\), and the subgroup \(\mathbb Z\subset \mathbb R\) is closed and discrete.
By assumption, the closed subgroup \(\Gamma\) may itself be gauged. Since \(\Gamma\) is discrete, gauging it introduces no gauge-covariant field strength, so the gauged theory still has \(\mathfrak a=0\).
On the other hand, its classifying space is
\begin{equation}
\mathcal A = \mathrm B\Gamma = \mathrm B\mathbb Z \simeq \mathbb S^1.
\end{equation}
But then
\begin{equation}
\pi_1(\mathcal A)\otimes\mathbb Q = \pi_1(\mathbb S^1)\otimes\mathbb Q = \mathbb Z\otimes\mathbb Q = \mathbb Q\ne 0,
\end{equation}
so \(\mathfrak l(\mathcal A)\neq 0\), contradicting the requirement \(\mathfrak a=0\). This contradiction shows that \(G\) cannot be noncompact.

It is important here that the subgroup \(\Gamma\) be closed. For instance, although \(\operatorname U(1)\) contains infinite cyclic subgroups generated by irrational rotations,
\begin{equation}
\left\{1,\exp(\pm2\pi\mathrm i\theta),\exp(\pm4\pi\mathrm i\theta),\dotsc\right\}\subset \operatorname U(1),
\qquad
\theta\notin \mathbb Q,
\end{equation}
these are dense in \(\operatorname U(1)\) and hence not closed.

\paragraph{Chevalley--Eilenberg algebra must capture the topology of the gauge group.}
We next argue more directly that a noncompact gauge symmetry is forbidden because the \(L_\infty\)-algebra \(\mathfrak a\) can only capture the correct topology of gauge bundles when the gauge group is compact. (Note that this argument is stronger than the previous one since, if a gaugeable noncompact group were consistent, one could gauge it to obtain a noncompact gauge group.)

Consider a gauge theory with gauge group \(G\) with Lie algebra \(\mathfrak g\) and the $L_\infty$-algebra \(\mathfrak{a}\) of invariants. Then \(\mathcal A\) will (among others) have to contain a factor of \(\mathrm BG\), corresponding to a principal \(G\)-bundle.
On the other hand, \(\mathfrak a\) will contain a factor \(\mathfrak{inv}(\mathfrak g)\).
For \cref{post:charge-quantisation} to hold, we should have \(\mathfrak l(\mathrm BG)\otimes_{\mathbb Q}\mathbb R = 
\mathfrak{inv}(\mathfrak g)\); however, if \(G\) admits an invariant positive-definite metric, then this only holds if \(G\) is compact.

For instance, consider \(G=\operatorname U(1)\). Then the rational homotopy type of \(\mathrm BG\simeq\mathbb{CP}^\infty\) is given by the Sullivan model \(\mathbb Q[x]\) with $|x|=2$ and \(\mathrm dx=0\), so that the Whitehead Lie algebra \(\mathfrak l(\mathrm BG)\otimes_\mathbb{Q}\mathbb R=\mathbb R[2]\) coincides with \(\mathfrak{a}\).
On the other hand, consider \(G=\mathbb R\). Then \(\mathrm B\mathbb R\simeq*\) is contractible (i.e.\ there is only one \(\mathbb R\)-bundle on a paracompact space up to topological isomorphism). Hence, a putative theory with gauge group \(\mathbb R\) violates \cref{post:charge-quantisation}.

These arguments lead us to the following claim.

\begin{constraint}
    Gauge groups and 't Hooft-anomaly-free connected Lie global symmetries must be compact.
\end{constraint}

An analogous conclusion is expected for higher-form symmetries: within the present charge-quantisation framework, compact groups such as \(\operatorname U(1)\) arise naturally, whereas noncompact groups such as \(\mathbb R\) lead to the same obstruction as in the ordinary Abelian gauge-field case. We leave a systematic treatment of this higher-form version for future work.

\subsection{Constraints on one-form field strengths}
There are algebras \(\mathfrak a\) for which there is no topological space \(\mathcal A\) such that \(\mathfrak l(\mathcal A)=\mathfrak a\). This rules out certain putative field theories.

Concretely, consider a theory with a family of invariant one-form field strengths \(F^1,\dotsc,F^n\in\Omega^1(M)\) that obey Bianchi identities
\begin{equation}\label{eq:one-form-bianchi}
    \mathrm dF^i + \frac12f^i_{jk}F^j\wedge F^k = 0.
\end{equation}
Consistency then requires that
\begin{equation}
    0 = \mathrm d(\mathrm dF^i) \propto f^i_{j[k}f^j_{lm]} F^k\wedge F^l\wedge F^m,
\end{equation}
so that \(f^i_{jk}\) form the structure constants of a Lie algebra \(\mathfrak a\), and (for the corresponding basis \(t_1,\dotsc,t_n\in\mathfrak a\)) we may define
\begin{equation}
    F = F^i t_i \in\Omega^1(M;\mathfrak a).
\end{equation}
Now, Sati--Schreiber charge quantisation implies that \(\mathfrak a\) must be nilpotent; in particular, it cannot be simple.

On the one hand, cases where \(\mathfrak a\) is nilpotent are clearly allowed, since in that case, in some suitable basis, we have
\begin{equation}
    \mathrm dF^1 = 0,\quad\dotsc,\quad
    \mathrm dF^i = \sum_{j,k<i} -f^i_{jk}F^j\wedge F_k,
\end{equation}
so that the first \(F^1\) is simply an Abelian field strength, and each \(F^i\) is defined on a background of \(F^j\) for \(j<i\).

On the other hand, such an ordering is \emph{not} possible for simple \(\mathfrak a\). In this case, the theory may be viewed as follows. Consider an \(\mathfrak h\)-valued gauge theory with potential given in a local patch \(U\) by \(A\in\Omega^1(U;\mathfrak h)\), constrained such that its field strength \(G=\mathrm dA+\frac12[A,A]=0\) always vanishes (for example, Chern--Simons theory or the \(BF\) model). Here, \(A\) is not a physical field and gauge-transforms. Suppose that one tries to reinterpret the theory by getting rid of gauge transformations, so that \(A\) is gauge-invariant and thus putatively physical; then the condition \(G=0\) may be interpreted as a \emph{Bianchi identity} on \(A\). In that case, after renaming \(A\) to \(F\) and \(\mathfrak h\) to \(\mathfrak h\), we arrive at \eqref{eq:one-form-bianchi} for a simple Lie algebra \(\mathfrak a\). This ill-fated manœuvre should not work, and the Sati--Schreiber charge-quantisation postulate rules it out.

Thus, we have the following constraint.

\begin{constraint}\label{conj:one-form-field-strengths}
Only nilpotent one-form flux algebras exist in a field theory.
\end{constraint}

\subsection{Bianchi identities and curvatures can only contain rational coefficients up to homotopy}
In \cref{post:charge-quantisation}, we have explicitly required that the \(L_\infty\)-algebras \(\mathfrak h\), \(\mathfrak w\), and \(\mathfrak a\) be defined over the rational numbers rather than the real numbers. However, this is far from immediate from the underlying quantum field theory: the various Bianchi identities that occur in \(\mathfrak w\) naturally come with real coefficients, and at the level of set-theoretic cardinality, most real \(L_\infty\)-algebras \(\mathfrak g_{\mathbb R}\) cannot be realised as \(\mathfrak g_{\mathbb R} = \mathfrak g_{\mathbb Q}\otimes_{\mathbb Q}\mathbb R\) for a rational \(L_\infty\)-algebra \(\mathfrak g_{\mathbb Q}\). Nevertheless, it is easy to see that this in fact holds for quantum field theories that naturally arise in practice. For instance, in Yang--Mills theory, the Lie algebra \(\mathfrak g\) is always reductive (a direct sum of Abelian and semisimple parts) and can be realised over the rational numbers, unlike more general Lie algebras.\footnote{For instance, the Bianchi classification of real three-dimensional Lie algebras includes two continuous families, so that the set of isomorphism classes of three-dimensional real Lie algebras has size \(2^{\aleph_0}\), whereas the set of isomorphism classes of three-dimensional rational Lie algebras has size \(\aleph_0\).}

Thus, we have the following constraint.
\begin{constraint}
The \(L_\infty\)-algebra \(\mathfrak a\) of curvatures and their Bianchi identities must be equivalent to an \(L_\infty\)-algebra defined over rational numbers up to homotopy.
\end{constraint}

\subsection{No generalised global symmetries and defect charges in quantum gravity}
The charge-quantisation law above suggests a homotopical form of the swampland no-global-symmetry \cite{Banks:1988yz} and completeness conjectures \cite{Palti:2019pca} in quantum gravity.
Since invertible Abelian topological generalised symmetries arising from the \(\mathcal A\)-valued structure are classified by \(\operatorname H^\bullet(\mathcal A; \operatorname U(1))=\hom(\operatorname H_\bullet(\mathcal A),\operatorname U(1))\), the no-global-symmetry conjecture implies that the space \(\mathcal A\) should have trivial reduced homology groups for a consistent theory of quantum gravity, i.e.\ \(\mathcal A\) should be an acyclic space. This is weaker than \(\mathcal A\) being contractible.
Furthermore, the completeness conjecture implies that all possible brane charges should already appear in the theory and hence be dynamical charges in the sense of \cref{ssec:brane};
in particular, since all charges are dynamical, there are no defect conserved charges, which would be classified by the homotopy groups \(\pi_k(\mathcal A)\) according to \cref{post:branes}. Therefore, the completeness conjecture implies that the homotopy groups of \(\mathcal A\) must be trivial in a consistent theory of quantum gravity.
Therefore, we have the following constraint.

\begin{constraint}\label{conj:no-generalised-global-symmetry}
In a consistent theory of quantum gravity, the classifying space \(\mathcal A\) should be contractible.
Equivalently, there are no non-trivial invertible Abelian generalised symmetries nor non-trivial defect charges in the \(\mathcal A\)-structure.
\end{constraint}

\section{Non-linear sigma models}\label{sec:nlsm}
As a simple example of the charge-quantisation postulate, consider a scalar field \(\phi\colon M\to X\) taking values in a smooth manifold \(X\) with action
\begin{equation}\label{eq:nlsm-action}
    S = \int_M\mathrm d^dx\,\sqrt{|\det g|}\left(-\frac12(\partial\phi)^2 - V(\phi)\right).
\end{equation}
In that case, on a Cauchy surface \(\Sigma\subset M\) in spacetime \(M\), the space of kinematically allowed states on \(\Sigma\), consisting of field configurations \(\phi\colon\Sigma\to X\) and their conjugate momenta, is the symplectic manifold
\begin{equation}
    \mathrm T^*\mathcal C^\infty(\Sigma,X),
\end{equation}
where the fibres of the cotangent bundle correspond to the momentum conjugate to \(\phi\). This space is homotopy-equivalent to \(\mathcal C^0(\Sigma,X)\), so that the space of classical states is classified by (the homotopy type of) the space \(X\); that is, the classifying space of the non-linear sigma model \eqref{eq:nlsm-action} is \(\mathcal A=X\).

\paragraph{Defect brane charges.}
According to \cref{post:branes}, the defect \(p\)-brane charges of the non-linear sigma model are classified by the homotopy group \(\pi_{d-p-2}(X)\). For the non-linear sigma models, these are the charges of skyrmions classified by the homotopy groups of the target space.

\paragraph{Higher-form symmetries.}
According to \cref{post:higher-form-symmetry}, the \(k\)-form symmetries of the non-linear sigma model correspond to the Pontryagin dual group of the \((d-k-1)\)\textsuperscript{st} homology group \(\operatorname H_{d-k-1}(X)\). Indeed, let \(\alpha\in\operatorname H^{d-k-1}(X;\operatorname U(1))\) be a cohomology class with \(\operatorname U(1)\) coefficients. Let \(\Sigma\subset M\) be a closed \((d-k-1)\)-dimensional submanifold. Then the topological operator \(\mathcal O_\alpha(\Sigma)\) corresponding to this \(k\)-form symmetry then simply weighs the path integral over different topological sectors with coefficients given by \(\int_M\phi^*\alpha\) as discussed in \eqref{eq:path-integral-operator}.

\section{Abelian (higher) gauge theory}\label{sec:abelian}
Consider Abelian \(p\)-form electrodynamics in \(d\) spacetime dimensions with a \(p\)-form potential valued in \(\operatorname U(1)\) and a dual \((d-p-2)\)-form potential valued in \(\operatorname U(1)\). Let the electric and magnetic gauge Lie algebras be \(\mathfrak g\otimes_{\mathbb Q}\mathbb R\) and \(\tilde{\mathfrak g}\otimes_{\mathbb Q}\mathbb R\), where both \(\mathfrak g\) and \(\tilde{\mathfrak g}\) are one-dimensional rational Lie algebras (i.e.\ isomorphic to \(\mathbb Q\)); we will distinguish between them, however, for clarity.
In this case, the gauge \(L_\infty\)-algebra \(\mathfrak h\) is simply the graded Abelian Lie algebra
\begin{align}
    \mathfrak h &= \mathfrak g[p-1]\oplus\tilde{\mathfrak g}[d-p-3],
\end{align}
and the corresponding inner-automorphism algebra \(\mathfrak w\) and algebra of invariants \(\mathfrak a\) are simply
\begin{equation}
    (\mathfrak h\to\mathfrak w\to\mathfrak a)
    =\left(
    \begin{tikzcd}[ampersand replacement=\&, row sep=0pt]
        \mathfrak g[p-1] \ar{r}{\operatorname{id}} \& \mathfrak g[p-1] \\
        \&\oplus \\
        \oplus \& \mathfrak g[p] \ar{r}{\operatorname{id}} \& \mathfrak g[p] \\
        \& \oplus \\
        \tilde{\mathfrak g}[d-p-3] \ar{r}{\operatorname{id}} \& \tilde{\mathfrak g}[d-p-3] \& \oplus\\
        \& \oplus \\
        \& \tilde{\mathfrak g}[d-p-2]\ar{r}{\operatorname{id}} \& \tilde{\mathfrak g}[d-p-2]
    \end{tikzcd}
    \right).
\end{equation}
We may integrate \(\mathfrak a\) into
\begin{equation}
    \mathcal A = \mathrm B^pU(1)\times\mathrm B^{d-p-2}\tilde U(1).
\end{equation}

\paragraph{Defect brane charges.}
The homotopy groups of \(\mathrm B^pU(1)\) are given by
\begin{equation}
    \pi_k(\mathrm B^pU(1)) = \begin{cases}
        \pi_{k-p}(U(1)) & \text{if \(k-p\ge1\)} \\
        U(1) & \text{if \(k-p=1\)} \\
        0 & \text{if \(k-p\le0\)},
    \end{cases}
\end{equation}
and similarly for \(\mathrm B^{d-p-2}\tilde U(1)\).
Since the homotopy groups of \(\operatorname U(1)\) are given by
\begin{equation}
    \pi_k(\operatorname U(1)) = \begin{cases}
        \mathbb Z &\text{if \(k=1\)}\\
        0 & \text{otherwise},
    \end{cases}
\end{equation}
its deloopings have the homotopy groups
\begin{equation}
    \pi_k(\mathrm B^p\operatorname U(1)) = \begin{cases}
        \mathbb Z &\text{if \(k=p+1\)}\\
        0 & \text{otherwise}.
    \end{cases}
\end{equation}
We thus we see that we obtain \(q\)-brane charges for \(d-q-2 \in \{p, d-p-2\}\), that is, \(q\in\{p,d-p-2\}\), corresponding to electrically and magnetically charged \(p\)- and \((d-p-2)\)-branes. In particular, when \(p=d-p-2\), then we obtain dyonic \(p\)-branes.

\paragraph{Higher-form symmetries.}
For concreteness, let us now specialise to \(p=1\) and \(d=4\), so that the space \(\mathcal A\) is
\begin{equation}
    \mathcal A = \mathrm{BU}(1)\times\mathrm{BU}(1)
    \simeq
    \mathbb{CP}^\infty\times\mathbb{CP}^\infty,
\end{equation}
that is, the square of the infinite-dimensional complex projective space.

The homology groups of \(\mathcal A\) in degrees \(\le4\) are
\begin{align}
    \operatorname H_0(\mathcal A) &= \mathbb Z,&
    \operatorname H_1(\mathcal A) &= 0,&
    \operatorname H_2(\mathcal A) &= \mathbb Z^2,&
    \operatorname H_3(\mathcal A) &= 0,&
    \operatorname H_4(\mathcal A) &= \mathbb Z^3,
\end{align}
so that the cohomology groups of \(\mathcal A\) with coefficients in \(\operatorname U(1)\) are the corresponding Pontryagin dual groups, namely
\begin{equation}
\begin{aligned}
    \operatorname H^0(\mathcal A;\operatorname U(1)) &= \operatorname U(1),&
    \operatorname H^1(\mathcal A;\operatorname U(1)) &= 0,\\
    \operatorname H^2(\mathcal A;\operatorname U(1)) &= \operatorname U(1)^2,&
    \operatorname H^3(\mathcal A;\operatorname U(1)) &= 0,&
    \operatorname H^4(\mathcal A;\operatorname U(1)) &= \operatorname U(1)^3.
\end{aligned}
\end{equation}
According to \cref{post:higher-form-symmetry}, the cohomology group \(\operatorname H^2(\mathcal A;\operatorname U(1))\) corresponds to one-form symmetries, which is the direct product of the \(\operatorname U(1)\) magnetic and \(\operatorname U(1)\) electric one-form symmetries, whose Noether currents are the magnetic and electric field strengths \(\mathrm dA\) and \(\mathrm d\tilde A=\star\mathrm dA\) respectively, in accordance with expectations.

In addition, there exists a trivial three-form symmetry corresponding to \(\operatorname H^0(\mathcal A;\operatorname U(1))\), whose Noether current is simply the identity local operator \(1\). Finally, there are three \(\operatorname U(1)\)-valued \((-1)\)-form symmetries corresponding to  \(\operatorname H^4(\mathcal A;\operatorname U(1))\), whose Noether currents are \(F\wedge F\) and \(\tilde F\wedge\tilde F\) (the electric and magnetic second Chern numbers) and \(F\wedge\tilde F\) (the Lagrangian density).

\subsection{Maxwell theory with sources}\label{ssec:maxwell-with-source}
Next, we examine what happens when we explicitly include matter fields
with electric charges \(m\mathbb Z\) and magnetic charges \(n\mathbb Z\).
For simplicity and definiteness, we continue specialise to four-dimensional Maxwell theory, i.e.\ \(d=4\) and \(p=1\).

The Bianchi identities are now
\begin{align}
    \mathrm d\tilde F &= J_\mathrm e,&
    \mathrm dJ_\mathrm e &= 0,&
    \mathrm dF &= J_\mathrm m,&    
    \mathrm dJ_\mathrm m&=0,
\end{align}
where \(F\) and \(\tilde F\) are the two dual two-form field strengths and \(J_\mathrm e\) and \(J_\mathrm m\) are the electric and magnetic currents respectively.
Hence the \(L_\infty\)-algebra \(\mathfrak a\) of Bianchi identities has trivial cohomology, unlike the case without matter fields.
As a consequence, \(\mathcal A\) only has torsion homotopy groups.
Recalling (see e.g.\ \cite[Example~3.4]{Bhardwaj:2023kri}) that the \(BF\) model in the deep infrared has \(\mathbb Z_m\)- and \(\mathbb Z_n\)-valued two-bundles, we have
\begin{equation}
    \mathcal A = \mathrm B^2\mathbb Z_m \times\mathrm B^2\mathbb Z_n.
\end{equation}

\paragraph{Defect brane charges.}
The homotopy groups of \(\mathcal A\) are trivial except for \(\pi_2\), where
\begin{equation}
    \pi_2(\mathcal A) = \mathbb Z_m\times\mathbb Z_n.
\end{equation}
By \cref{post:branes}, the possible charges of defect line operators (zero-branes) are valued in \(\mathbb Z_m\times\mathbb Z_n\),
which is reduced from the sourceless case \(\pi_2(\mathrm{BU}(1)\times\mathrm{BU}(1))=\mathbb Z\times\mathbb Z\), since multiples of \(m\) (for electric charges) and \(n\) (for magnetic charges) have been gauged away.

\paragraph{Higher-form symmetries.}
According to the Hurewicz theorem, which relates homotopy groups and homology groups,
the lowest-degree nontrivial homology group of \(\mathcal A\) is
\begin{equation}\operatorname H_2(\mathcal A)=\pi_2(\mathcal A) = \mathbb Z_m\times\mathbb Z_n,\end{equation}
which corresponds to the well-known breaking of the electric and magnetic one-form symmetries into cyclic subgroups. 
In addition, we have a nontrivial \(\operatorname H_4(\mathcal A)\), which corresponds to a zero-form symmetry given by characteristic classes of the \(\mathbb Z_m\times\mathbb Z_n\)-valued two-bundle.

\section{Yang--Mills theory}\label{sec:yang-mills}
Let \( G\) be a compact Lie group. Let us consider a gauge theory with gauge group \(G\) on a spacetime manifold \(M\). We assume for simplicity that \(G\) is connected, so that\footnote{\Cref{eq:lie-of-centre-is-centre-of-lie} fails for disconnected \(G\). For instance, if \(G=\operatorname O(2)\), since \(\mathfrak{lie}(\operatorname Z(G))=0\) but \(\mathfrak z(\mathfrak{lie}(G)) = \mathbb R\).}
\begin{equation}\label{eq:lie-of-centre-is-centre-of-lie}
    \mathfrak{lie}(\operatorname Z(G)) = \mathfrak z(\mathfrak{lie}(G)),
\end{equation}
where \(\operatorname Z(-)\) denotes the centre of a group and \(\mathfrak z(-)\) denotes the centre of a Lie algebra.
The Lie algebra \(\mathfrak g=\mathfrak{lie}(G)\) is a direct sum of an Abelian Lie algebra and a compact semisimple Lie algebra; in particular, it can be canonically defined as a Lie algebra over the rational numbers \(\mathbb Q\).

Since we are dealing with an ordinary gauge theory, no adjustments are necessary, and we may work with the canonical (unadjusted) inner-derivation algebra \(\mathfrak{inn}({\mathfrak g})\) and the algebra of invariant polynomials \(\mathfrak{inv}({\mathfrak g})\), which fit into the short exact sequence
\begin{equation}
    {\mathfrak g}\to\mathfrak{inn}({\mathfrak g})\to\mathfrak{inv}({\mathfrak g}),
\end{equation}
which lifts to the sequence of topological spaces
\begin{equation}
     G\to\mathrm E G\to\mathrm B G.
\end{equation}
One is therefore tempted to consider \(\mathcal A\overset?=\mathrm B G\), which classifies principal \( G\)-bundles. The fact that this does not suffice can be seen from two directions.
\begin{itemize}
    \item The space \(\mathrm B G\) does not manifestly display electric--magnetic duality in \(d=4\) (and for the Abelian component of \( G\) in other dimensions).
    \item The classifying space \(\mathrm B G\) only captures the kinematics of the connection \(A\) of a principal \( G\)-bundle. Therefore it knows about what may be termed off-shell (or magnetic) symmetries in the sense that, if we consider \( G=\operatorname U(1)\) for simplicity, then the fact that the curvature \(F=\mathrm dA\) defines a magnetic \((d-3)\)-form symmetry holds independent of the equations of motion.\footnote{i.e.\ the descent chain terminates, from the perspective of \cite{Borsten:2025pbx}} On the other hand, it does not know about the electric one-form symmetry given by \(\star F\), since the fact that \(\mathrm d(\star F)=0\) is a consequence of the equations of motion.
\end{itemize}
Of course, these two reasons are two sides of the same coin in the sense that electric--magnetic duality interchanges the Bianchi identity with the equations of motion and hence exchanges off-shell magnetic symmetries with on-shell electric symmetries. Hence, one needs to `double' \(\mathrm B G\), treating the electric and magnetic principal bundles democratically. Since electric--magnetic duality depends on whether the number of spacetime dimension is \(d=4\) or \(d>4\), we treat the two cases separately.

\subsection{Four-dimensional Yang--Mills theory}
In four dimensional Yang--Mills theory with gauge group \( G\), the Wilson lines are valued in representations of \(G\) and hence represent holonomies of a connection on a principal \( G\)-bundle on spacetime. Dually, the 't~Hooft lines are valued in representations of \(\langlands{ G}\) \cite{Goddard:1976qe} and should be interpreted as holonomies of a connection on a principal \(\langlands{ G}\)-bundle on spacetime.
In particular, if the matter content is such that one obtains \(\mathcal N=4\) supersymmetric Yang--Mills theory,
then this theory enjoys Montonen--Olive S-duality \cite{Montonen:1977sn},
such that under the generator \(\mathsf S\) of \(\operatorname{PSL}(2;\mathbb Z)\), 
a \(G\)-valued supersymmetric Yang--Mills theory is dual to a \(\langlands{ G}\)-valued supersymmetric Yang--Mills theory \cite{Kapustin:2006pk} (at least when the theta angles are set to zero; see \cite{Aharony:2013hda} for the case with theta angles).
Therefore, on a Cauchy surface \(\Sigma\) there exist \emph{two} principal bundles: a principal \( G\)-bundle and a principal \(\langlands{ G}\)-bundle, and the kinematic data on \(\Sigma\) consist of two principal bundles \(P_{ G}\) and \(P_{\langlands{ G}}\) along with connections \(A_i\), \(\langlands A_i\) and associated canonical momenta that satisfy the Gauss laws and other constraints.
Thus, on a Cauchy surface \(X\), the physical data contain the principal bundles \(P_{ G}\) and \(P_{\langlands{ G}}\) (in addition to the connections on them and their conjugate momenta); this means that we have a projection\footnote{
    More precisely, to take into account the connections,
    we should be looking at the \emph{smooth} classifying stack \(\mathrm B_\nabla( G\times\langlands{ G})\) that classifies both the principal bundle as well as the connection, and there is a forgetful map \(\mathrm B_\nabla( G\times\langlands{ G})\to\mathrm B( G\times\langlands{ G})\) that forgets the connection \cite{Schreiber:2013pra}.
    However, \(\mathrm B_\nabla( G\times\langlands{ G})\) is not a homotopy type but a smooth stack (or simplicial manifold). The simplicial realisation of \(\mathrm B_\nabla( G\times\langlands{ G})\) has the same homotopy type as \(\mathrm B( G\times\langlands{ G})\).
}
\begin{equation}
    \mathcal A\to \mathrm B G\times\mathrm B\langlands{ G}.
\end{equation}
For the purposes of this section, we ignore the finer (smooth) structure of \(\mathcal A\) and simply regard
\begin{equation}
    \mathcal A = \mathrm BH=\mathrm B G \times \mathrm B\langlands{ G},
\end{equation}
where \(\langlands{ G}\) is the Langlands dual group of \( G\), and where we have defined the duality-invariant gauge group
\begin{equation}
    H \coloneqq  G\times\langlands{ G}.
\end{equation}

The S-duality-invariant gauge algebra then is:
\begin{equation}
    \mathfrak h={\mathfrak g}\oplus\langlands{{\mathfrak g}}.
\end{equation}
The inner-derivation algebra \(\mathfrak w\) is the ordinary (unadjusted) inner-derivation algebra \(\mathfrak{inn}(\mathfrak h)\), and similarly \(\mathfrak a=\mathfrak{inv}(\mathfrak h)\) is the ordinary algebra of invariant polynomials; the short exact sequence
\begin{equation}
    \mathfrak h\to\mathfrak w\to\mathfrak a
\end{equation}
lifts to
\begin{equation}
    H\to\mathrm EH\to\mathrm BH.
\end{equation}

\paragraph{Defect brane charges.}
The charges of the defect \(p\)-branes of four-dimensional Yang--Mills theory are therefore classified by the generators of the homotopy groups
\begin{equation}
    \pi_{2-p}(\mathrm BH)
    =\begin{cases}
        \pi_{1-p}(H) & \text{if \(p\le0\)}\\
        1 & \text{if \(p\ge1\)}.
    \end{cases}
\end{equation}
However, for \(p=-1\), the second homotopy group \(\pi_2(H)\) of any Lie group is trivial.
The cases \(p<-1\) are not allowed by dimensionality.
Therefore, the only non-trivial case is for \(p=0\), that is, zero-branes (or line operators in spacetime), whose charges are
\begin{equation}\pi_1(H) = \pi_1( G)\times\pi_1(\langlands{ G})
    =\pi_1( G)\times\operatorname Z( G).
\end{equation}
These are pure torsion and correspond to the `classes labelled by a pair' that classify dyonic line operators in \cite{Aharony:2013hda}.

\paragraph{Higher-form symmetries.}
Next, we examine the higher-form symmetries. According to \cref{post:higher-form-symmetry}, an invertible Abelian \(k\)-form symmetry should correspond to the Pontryagin dual of \(\operatorname H_{3-k}(\mathrm BH)\), which is given by the Künneth theorem as
\begin{equation}
    \operatorname H_{3-k}(\mathrm BH)
    =
    \bigoplus_{\mathclap{i+j=3-k}}\operatorname H_i(\mathrm BG)\otimes\operatorname H_j(\mathrm B\langlands G)
    \oplus
    \bigoplus_{\mathclap{i+j=2-k}}
    \operatorname{Tor}_1^{\mathbb Z}\mleft(\operatorname H_i(\mathrm BG),\operatorname H_j(\mathrm B\langlands G)\mright)
    .
\end{equation}
For a compact connected Lie group (i.e. for both $G$ and $\langlands G$), recall that
\begin{align}
\operatorname H_0(\mathrm BG)&=\mathbb Z,&
\operatorname H_1(\mathrm BG)&=0,&
\operatorname H_2(\mathrm BG)&=\pi_1(G),&
\operatorname H_3(\mathrm BG)&=0,&
\operatorname H_4(\mathrm BG)&=\mathbb Z.
\end{align}
Hence, we obtain
\begin{align}
    \operatorname H_0(\mathrm BH) &= \mathbb Z,&
    \operatorname H_1(\mathrm BH) &= 0,\\
    \operatorname H_2(\mathrm BH) &= \pi_1(G)\oplus\operatorname Z(G),&
    \operatorname H_3(\mathrm BH) &= 0,&
    \operatorname H_4(\mathrm BH) &= \pi_1(G)\otimes\operatorname Z(G)\oplus\mathbb Z^2.\notag
\end{align}
Note that, for degree reasons, the \(\operatorname{Tor}\) term does not contribute in four spacetime dimensions. That is, the invertible generalised symmetries are given by the corresponding Pontryagin dual groups,
\begin{equation}
\begin{aligned}
    \operatorname H^0(\mathrm BH;\operatorname U(1)) &= \operatorname U(1),&
    \operatorname H^1(\mathrm BH;\operatorname U(1)) &= 0,\\
    \operatorname H^2(\mathrm BH;\operatorname U(1)) &= \pi_1(G)\oplus\operatorname Z(G),&
    \operatorname H^3(\mathrm BH;\operatorname U(1)) &= 0,\\
    \operatorname H^4(\mathrm BH;\operatorname U(1)) &= \pi_1(G)\otimes\operatorname Z(G)\oplus\operatorname U(1)^2.
\end{aligned}
\end{equation}

The three-form symmetry given by \(\operatorname H^0(\mathrm BH;\operatorname U(1))\) is the trivial symmetry whose Noether current is the identity operator \(1\). The one-form symmetry given by \(\operatorname H^2(\mathrm BH;\operatorname U(1))=\pi_1(G)\oplus\operatorname Z(G)\) corresponds to the magnetic and electric centre symmetries respectively; the electric one-form centre symmetry, in particular, plays a key role in confinement \cite{Polyakov:1975rs,Polyakov:1976fu,tHooft:1977nqb}, as reviewed in \cite{Holland:2000uj,Ogilvie:2012is}.

Finally, one has the \((-1)\)-form symmetry given by \(\operatorname H^4(\mathrm BH;\operatorname U(1)) = \pi_1(G)\otimes\operatorname Z(G)\oplus\operatorname U(1)^2\). The torsion-free part \(\operatorname U(1)^2\) is captured by rational homotopy theory (that is, the algebra \(\mathfrak a\)) and correspond to invariant polynomials; in four dimensions, due to degree reasons we only obtain the \((-1)\)-form symmetry given by the second Chern class \(\operatorname{tr}(F\wedge F)\) of \(P_{ G}\) (where \(F\) is the field strength of the connection). Dually, we have the magnetic \(\operatorname{tr}(\langlands F\wedge\langlands F)\), which is the second Chern class of \(P_{\langlands{ G}}\) (where \(\langlands F\) is the dual field strength).
In the case of \(\mathcal N=4\) supersymmetric Yang--Mills theory, these operators are related to the two \(\theta\)-angles related by S-duality, which are independent in the following sense: under Montonen--Olive duality, the coupling constant
\begin{equation}
    \tau=\frac\theta{2\pi}+\frac{4\pi\mathrm i}{g^2}
\end{equation}
transforms as
\begin{equation}
    \tau\mapsto -\frac1{n_{ G}\tau}
\end{equation}
with \(n_{ G}\) the dual Coxeter number of \( G\), so that in particular the dual \(\theta\)-angle is
\begin{equation}
    \langlands\theta = 2\pi\Re\mleft( -\frac1{n_{ G}\tau}\mright),
\end{equation}
which mixes \(\theta\) and \(g\) and is not a function of \(\theta\) alone. Hence \(\operatorname{tr}(F\wedge F)\) and the dual \(\operatorname{tr}(\langlands F\wedge\langlands F)\) (with \(\langlands F\) the dual field strength) are, in general, distinct operators.

The torsion \((-1)\)-form symmetry, valued in \(\pi_1(G)\otimes\operatorname Z(G)\), is the tensor product of the magnetic and electric centre symmetries. This operator, when inserted in the path integral for a spacetime \(M\) with the bundles \(P_G\) and \(P_{\langlands G}\), takes the value
\begin{equation}
    \int_M c(P_G) \smile c(P_{\langlands G}),
\end{equation}
where \(c(P_G)\in\operatorname H^2(M)\) is the cohomology class such that the electric one-form symmetry operator is \(\int_{\Sigma_2} c(P_G)\), while \(c(P_G)\in\operatorname H^2(P_{\langlands G})\) is the analogous cohomology class for the magnetic one-form symmetry.

\subsection{Higher-dimensional Yang--Mills theory}
Consider a \(G\)-valued Yang--Mills theory in \(d>4\) dimensions.
In \(d>4\), non-Abelian electric--magnetic duality no longer holds because a \(p\)-form gauge field (i.e.\ concentrated in degree \(p\)) cannot be non-Abelian, so the previous construction of \(\mathcal A\) for \(d=4\) no longer applies.
Since the data of Yang--Mills theory include a principal \(G\)-bundle, there must be a projection \(\mathcal A\twoheadrightarrow\mathrm BG\). Just taking \(\mathcal A=\mathrm BG\) is, however, inconsistent with the existence of electric--magnetic duality in the Abelian case when \(G=\operatorname U(1)\) and furthermore, in the non-Abelian case, fails to reproduce the correct electric symmetries and line-operator charges. The simplest possibility consistent with the Abelian case is
\begin{equation}
    \mathcal A = \mathrm BG \times\mathrm B^{d-2}\widehat{\operatorname Z(G)},
\end{equation}
where \(\widehat{\mathrm Z(G)}\) is the Pontryagin dual of \(\operatorname Z(G)\). When \(G=\operatorname U(1)\), this specialises to
\begin{equation}
    \mathcal A = \mathrm{BU}(1) \times\mathrm B^{d-2}\hat{\mathbb Z}
    =\mathrm{BU}(1) \times\mathrm B^{d-3}\mathrm U(1)
\end{equation}
as expected. Note that, for \(G\) a compact connected simple Lie group, \(\operatorname Z(G)\) is a product of cyclic groups, so that it is isomorphic to its own Pontryagin dual; since the Abelian case has already been discussed in \cref{sec:abelian}, henceforth we assume \(G\) to be a compact connected simple Lie group.

\paragraph{External brane charges.}
According to \cref{post:branes}, the charges of defect \(p\)-branes of \(d\)-dimensional Yang--Mills theory are classified by the homotopy groups of \(\mathcal A\).
\begin{equation}
    \pi_{d-p-2}(\mathcal A)
    =\pi_{d-p-2}(\mathrm BG)
    \times
    \pi_{d-p-2}(\mathrm B^{d-2}\operatorname Z(G))
    =
    \pi_{d-p-3}(G)
    \times
    \pi_{-p}(\operatorname Z(G)).
\end{equation}
In particular, for \(p=0\), we obtain Wilson line operators labelled by \(\operatorname Z(G)\) (similar to the coarse classification of line operators in \cite{Aharony:2013hda}, except that we no longer have dyonic line operators), in addition to \(p\)-branes with (possibly torsion) magnetic charges given by a non-trivial \(G\)-bundle around the transverse directions.

\paragraph{Higher-form symmetries.}
According to \cref{post:higher-form-symmetry},
the \(k\)-form symmetries of \(d>4\)-dimensional Yang--Mills theory are classified by the Pontryagin dual groups of
\begin{equation}
    \operatorname H_{d-k-1}(\mathcal A)
    =
    \bigoplus_{\mathclap{i+j=d-k-1}}\operatorname H_i(\mathrm BG)\otimes\operatorname H_j(\mathrm B^{d-2}\mathrm Z(G))
    \oplus
    \bigoplus_{\mathclap{i+j=d-k-2}}
    \operatorname{Tor}_1^{\mathbb Z}\mleft(\operatorname H_i(\mathrm BG),\operatorname H_j(\mathrm B^{d-2}\mathrm Z(G))\mright)
    .
\end{equation}
Now, by the Hurewicz theorem,
\begin{equation}
\begin{aligned}
    \operatorname H_0(\mathrm B^{d-2}\mathrm Z(G))&=\mathbb Z,\\
    \operatorname H_i(\mathrm B^{d-2}\mathrm Z(G))&=0\qquad(1\le i\le d-3),\\
    \operatorname H_{d-2}(\mathrm B^{d-2}\mathrm Z(G))&=\operatorname Z(G).
\end{aligned}
\end{equation}
Hence, by the Künneth theorem,
\begin{equation}
    \operatorname H_i(\mathcal A) = \begin{cases}
        \operatorname H_i(\mathrm BG) & \text{if \(i\le d-3\)} \\
        \operatorname H_{d-2}(\mathrm BG)\oplus\operatorname Z(G) & \text{if \(i=d-2\)} \\
        \operatorname H_{d-1}(\mathrm BG)\oplus
        \operatorname H_{d-1}(\mathrm B^{d-2}\mathrm Z(G))
        & \text{if \(i=d-1\)} \\
        \operatorname H_d(\mathrm BG)\oplus
        \pi_1(G)\otimes\operatorname Z(G)
        \oplus
        \operatorname H_d(\mathrm B^{d-2}\mathrm Z(G)) & \text{if \(i=d\)}.
    \end{cases}
\end{equation}
The non-torsion parts of \(\operatorname H_\bullet(\mathrm BG)\) (corresponding to invariant polynomials of the field strength \(F\)) as well as the \(\operatorname Z(G)\) part of \(\operatorname H_{d-2}(\mathcal A)\) (corresponding to the \(\operatorname Z(G)\)-valued electric one-form symmetry) and the torsion \(\operatorname H_2(\mathcal A)=\operatorname H_2(\mathrm BG)=\pi_1(G)\) (corresponding to the \(\pi_1(G)\)-valued magnetic \((d-3)\)-form symmetry) are expected. In addition, other torsion parts of \(\operatorname H_\bullet(\mathrm BG)\), as well as higher cohomologies of \(\mathrm B^{d-2}\mathrm Z(G)\), produce higher analogues of the magnetic \((d-3)\)-form symmetry that do not seem to already appear in the literature.

\section{String theory with 16 supercharges and adjusted higher gauge theory}\label{sec:adjusted}
So far, we have looked at non-gravitational theories. For a gravitational theory, it is expected that there are no global symmetries, including higher-form symmetries.\footnote{Nevertheless, it has been argued that the centre of the Lorentz group should produce a gravitational one-form symmetry \cite{Cheung:2024ypq}.} This means that the classifying space \(\mathcal A\) (if it exists) should be contractible or, at least, have trivial homologies \(\operatorname H_i(\mathcal A)\) for \(i\le d\) for a \(d\)-dimensional gravitational theory.

We note that, under \cref{post:branes}, this does \emph{not} mean that there are no stable branes --- rather, it means that there are no gauge-invariant, localised, quantised, conserved brane charges \cite{Marolf:2000cb}. That is, string theories feature modified Bianchi identities, which means that there does not exist one unique notion of charge but rather several such notions  (Maxwell charge, Page charge, brane-source charge), each of which violate at least one of gauge invariance, locality, quantisation, or conservation \cite{Marolf:2000cb}. In this respect, the present analysis does not contract the usual K-theoretic analysis of brane charges \cite{Witten:1998cd,Witten:2000cn} (reviewed in \cite{Olsen:1999xx,Szabo:2002jv,Moore:2003vf,Manjarin:2004ij,Evslin:2006cj,Szabo:2008hx}), which would be given by \(\mathcal A\) being the K-theory spectrum \cite{Sati:2025vjw}.

In this section, we consider the brane charges and higher-form symmetries of Type~I string theory (or equivalently heterotic string theory with gauge group \(G=\operatorname{Spin}(32)/\mathbb Z_2\), and explain how the gauge structure bears out this expectation.
The gauge structure is given by adjusted higher gauge theory \cite[§4.5]{Borsten:2024gox}, whose gauge structure is given by Whitehead tower (cf.\ \cref{ssec:whitehead-tower}) of \(G\), starting with the string two-group \(\operatorname{String}(G)\).
In ten-dimensional theories with sixteen supercharges, the only other possible gauge group is \((\mathrm E_8\times\mathrm E_8)\rtimes\mathbb Z_2\) \cite{Kaidi:2023tqo}.\footnote{Other gauge Lie algebras are ruled out by anomaly cancellation \cite{Adams:2010zy}.} This case can be treated similarly except for the fact that the group is disconnected (i.e.\ there is discrete gauging). Let \(\mathfrak g=\mathfrak o(32)\) be the corresponding Lie algebra.

\subsection{The \(L_\infty\)-algebraic gauge structure of string theory with 16 supercharges}
The bosonic sector of the low-energy effective action of \(\mathcal N=(1,0)\) superstring theories in Einstein frame is \cite[eq.\ (2.216)]{Blumenhagen:2006ci}
\begin{multline}
    S[g,\phi,A,B] = \int\frac{\sqrt{|\det g|}}{2\kappa_{10}^2}\Bigg(\exp(-2\phi)\left(R+4(\partial\phi)^2\right)\\
    -\frac1{12}H_{\mu\nu\rho}H^{\mu\nu\rho}- \frac{\kappa_{10}^2}{2g_\mathrm{YM}^2}\operatorname{tr}(F_{\mu\nu}F^{\mu\nu}) + \dotsb\Bigg),
\end{multline}
where the ellipsis denotes higher-order terms.
In the above, \(F\) denotes the \(\mathfrak g\)-valued field strength of the Yang--Mills potential \(A\) 
and \(H\) the three-form field strength corresponding to the two-form Kalb--Ramond field \(B\). $H$ and its dual seven-form field strength \(\tilde H\) satisfy the Bianchi identities \cite[eq.\ (2.214)]{Blumenhagen:2006ci} \cite{Freed:2000ta,Sati:2008kz}
\begin{equation}\label{eq:type-I-bianchi}
\begin{aligned}
    \mathrm dH &= 2\pi\left(\operatorname{ch}_2(F)-\frac12p_1\right),\\
    \mathrm d\tilde H &= 2\pi\left(\operatorname{ch}_4(F)
    -\frac1{48}\operatorname{ch}_2(F)p_1
    +\frac1{64}p_1^2 -\frac1{48}p_2
    \right),
\end{aligned}
\end{equation}
where \(p_i\) is the Pontryagin class of the tangent bundle of degree \(4i\), and \(\operatorname{ch}_i\) is the degree \(i\) component of the Chern character \(\operatorname{ch}(F)=\operatorname{tr}\exp(F/2\pi\mathrm i)\).

For simplicity, let us work on a spacetime that is parallelisable, such as (any spacetime diffeomorphic to) a torus or flat space; in that case, the Pontryagin classes of the tangent bundle vanish.\footnote{More generally, if we wish to include the Pontryagin classes, then this means that instead of considering maps \(M\to\mathcal A\) from spacetime to a classifying space \(\mathcal A\), we should be considering sections of possibly non-trivial \(\mathcal A\)-bundles over \(M\), i.e.\ twisted homotopy classes, similar to the discussion \cite{Fiorenza:2019usl} in the context of Hypothesis H.}

Thus, the \(L_\infty\)-algebra \(\mathfrak h\) of potentials has the underlying graded vector space
\begin{equation}
    \mathfrak h = \mathfrak g\oplus\mathbb R[1]\oplus\mathbb R[5],
\end{equation}
spanned by the basis vectors \(\mathtt a_a\) (for \(\mathfrak g\), representing the Yang--Mills potential \(A\)), \(\mathtt b\) (for \(\mathbb R[1]\), representing the Kalb--Ramond two-form \(B\)), and \(\tilde{\mathtt b}\) (for \(\mathbb R[5]\), representing the dual Kalb--Ramond six-form \(\tilde B\)), with \(L_\infty\)-algebra brackets given by its Chevalley--Eilenberg algebra
\begin{align}
    \mathrm d\mathtt a^a &= \frac12f^a_{bc}\mathtt a^b\mathtt a^c,&
    \mathrm d\mathtt b &= \alpha_3(\mathtt a) = \frac16f_{abc}\mathtt a^a\mathtt a^b\mathtt a^c,&
    \mathrm d\tilde{\mathtt b} &= \alpha_7(\mathtt a),
\end{align}
where \(\alpha_3\) and \(\alpha_7\) are the unique (up to scaling) non-trivial cocycles of degrees three and seven in the Lie algebra cohomology of \(\mathfrak g=\mathfrak o(32)\). Here, \(a,b,c\) are adjoint indices for \(\mathfrak g\).
This \(L_\infty\)-algebra has been called the \emph{fivebrane algebra} in \cite{Sati:2008kz}; if \(\tilde{\mathtt b}\) is excised, then this reduces to the \emph{string algebra} of \(\mathfrak g\).

We encode the Bianchi identities \eqref{eq:type-I-bianchi} into the adjusted inner-derivation algebra \(\mathfrak w\) of \(\mathfrak h\), which is of the type that has been called the `Chern--Simons \(L_\infty\)-algebra' in \cite[Def.~23]{Sati:2008eg}; in the case where \(\tilde{\mathtt b}\) hass been excised, this has been described in detail in \cite{Saemann:2017rjm,Schmidt:2019pks} (cf.\ the review in \cite{Borsten:2021ljb}).
Concretely, \(\mathfrak w\) has underlying graded vector space
\begin{equation}
    \mathfrak w = \underbrace{\mathfrak g}_{\mathtt a_a}\oplus\underbrace{\mathfrak g[1]}_{\mathtt f_a}\oplus\underbrace{\mathbb R[1]}_{\mathtt b}\oplus\underbrace{\mathbb R[2]}_{\mathtt h}\oplus\underbrace{\mathbb R[5]}_{\tilde{\mathtt b}}\oplus\underbrace{\mathbb R[6]}_{\tilde{\mathtt h}},
\end{equation}
with basis vectors as indicated, and whose \(L_\infty\)-algebra brackets are given by its Chevalley--Eilenberg algebra as
\begin{equation}
\begin{aligned}\label{eq:adjusted-weil-bianchi}
    \mathrm d\mathtt a^i &= \mathtt f^i+\frac12f^i_{jk}\mathtt a^j\mathtt a^k, &
    \mathrm d\mathtt b &= \mathtt h + \operatorname{CS}_3(\mathtt a)/2,&
    \mathrm d\tilde{\mathtt b} &= \tilde{\mathtt h} + \operatorname{CS}_7(\mathtt a)/4!, \\
    \mathrm d\mathtt f^i &= f^i_{jk}\mathtt a^j\mathtt f^k, &
    \mathrm d\mathtt h &= \operatorname{tr}(\mathtt f^2)/2, &
    \mathrm d\mathtt h &= \operatorname{tr}(\mathtt f^4)/4!,
\end{aligned}
\end{equation}
where \(\operatorname{CS}_3\) and \(\operatorname{CS}_7\) are the Chern--Simons three- and seven-forms respectively.
In this case, \(H\) and \(\tilde H\) are gauge-invariant but not closed; their Bianchi identities cannot be written in terms of gauge-invariant field strengths, only gauge-covariant ones. The presence of the Chern--Simons terms means that the adjusted inner-derivation algebra is not the canonical one but instead involves a non-trivial adjustment given by the Bianchi identities \eqref{eq:type-I-bianchi}.

The adjusted algebra of invariants is then given by the quotient
\(\mathfrak a = \mathfrak w/\mathfrak g\).
Its underlying graded vector space is
\begin{equation}
    \mathfrak a = \underbrace{\mathfrak{inv}(\mathfrak g)}_{\mathtt p_1,\dotsc,\mathtt p_{16}}\oplus\underbrace{\mathbb R[2]}_{\mathtt h},
\end{equation}
where \(\mathfrak{inv}(\mathfrak g)\) is the graded vector space of invariant polynomials of \(\mathfrak g=\mathfrak o(32)\) of total dimension \(16\), spanned by \(\mathtt p_1,\dotsc,\mathtt p_{16}\), and \(\mathbb R[2]\) corresponds to \(\mathtt h\) (the three-form field strength); its \(L_\infty\)-algebraic brackets are given by its Chevalley--Eilenberg algebra as
\begin{align}
    \mathrm d\mathtt h &= \mathtt p_1,&
    \mathrm d\tilde{\mathtt h} &= \mathtt p_2,&
    \mathrm d\mathtt p_i &= 0,& 
\end{align}
where \(\mathtt p_1 = \operatorname{tr}(\mathtt f^2)/2\) and \(\mathtt p_2=\operatorname{tr}(\mathtt f^4)/4!\) corresponds to the two lowest invariant polynomials of \(\mathfrak g\).
In particular, \(\mathtt h\) and \(\mathtt p_1\) form a trivial pair and disappear from the cohomology, and the same holds for \(\tilde{\mathtt h}\) and \(\mathtt p_2\).

The higher \(\mathtt p_i\) have not been trivialised solely for degree reasons. That is, one can imagine, in addition to the usual field content of Type~I string theory, a tower of ten-form, 14-form, etc.\ potentials that would successively trivialise all the other \(\mathtt p_i\)s; these fields, of course, would not have any dynamical degrees of freedom since they are defined on ten-dimensional spacetime, so we can formally include them into \(\mathfrak h\), \(\mathfrak w\), and \(\mathfrak a\) without harm.\footnote{Indeed, it has been argued in the literature that it is natural to add a ten-form potential to \(\mathfrak h\), which results in what has been called the \emph{ninebrane algebra} \cite{Sati:2014yxa}, with speculations that it is connected to nine-branes in string and M-theory.} In that case, \(\mathfrak a\) has trivial cohomology and becomes quasi-isomorphic to the zero-dimensional \(L_\infty\)-algebra, as expected for a gravitational theory. We now assume this henceforth.

One way to think of this trivialisation is that, starting from the lowest degree, each non-nontrivial non-torsion homotopy group of \(\mathrm BG\) is trivialised by a corresponding potential or, equivalently, a BPS brane. That is, each non-torsion element of \(\pi_i(G)\) is trivialised by the existence of a BPS \((8-i)\)-brane.
This is equivalent to the rational Whitehead tower of \(\mathfrak g=\mathfrak o(32)\).

\subsection{The contractible classifying space}
We next construct the classifying space \(\mathcal A\) whose Whitehead algebra is \(\mathfrak a\).
On the one hand, since \(\mathfrak a\) is contractible, we may immediately conclude that \(\mathcal A\) is also contractible. On the other hand, it is instructive to see how this happens explicitly; we will see that each (BPS or non-BPS) brane trivialises successive homotopy groups of \(G=\operatorname{Spin}(32)/\mathbb Z_2\), leading to the Whitehead tower of \(G\).

Naïvely, due to the existence of the Yang--Mills connection valued in \(G=\operatorname{Spin}(32)/\mathbb Z_2\), one expects \(\mathcal A\) to be approximated by \(\mathrm BG\). However, \(\mathcal A\) differs from \(\mathrm BG\) at least in that the torsion components of the homotopy groups of \(\mathrm BG\) have been trivialised by BPS branes. The non-trivial low-degree homotopy groups of \(\mathrm BG\) are
\begin{align}
    \pi_{i+1}(\mathrm BG) = 
    \pi_i(G) = 
    \begin{cases}
        \mathbb Z_2 &\text{if \(i\equiv 0,1\pmod8\) and \(i\ne0\)}\\
        \mathbb Z &\text{if \(i\equiv 3,7\pmod8\)}\\
        0 &\text{otherwise}
    \end{cases}\qquad(i\le 30).
\end{align}
These are given (with suitable modifications for \(\pi_0(G)\) and \(\pi_1(G)\)) by the Bott periodicity of the homotopy groups of \(\operatorname O(32)\) in the stable region, which yields
\begin{equation}
    \pi_i(\operatorname O(n)) = \begin{cases}
        \mathbb Z_2 &\text{if \(i\equiv 0,1\pmod8\)}\\
        \mathbb Z &\text{if \(i\equiv 3,7\pmod8\)}\\
        0 &\text{otherwise}
    \end{cases}\qquad(i\le n-2).
\end{equation}
The two lowest non-torsion components \(\pi_4(\mathrm BG)\) and \(\pi_8(\mathrm BG)\) correspond to the D5-brane and D1-brane in Type~I string theory (or, equivalently under S-duality, the NS5-brane and the fundamental string in \(\operatorname{Spin}(32)/\mathbb Z_2\)-valued heterotic string theory).

It is natural to suppose, then, that the torsion homotopy groups of \(\mathrm BG\) are trivialised by torsionful non-BPS branes, so that \(\mathcal A\) becomes the colimit of the Whitehead tower of \(G\), which is contractible.
Indeed, it has been recently argued \cite{Kaidi:2023tqo} that every homotopy group element of \(G\) corresponds to some stable brane, which are necessarily non-BPS for torsion elements. For \(\pi_9(\mathrm B(\operatorname{Spin}(32)/\mathbb Z_2))\), this brane is the \(\mathbb Z_2\)-valued D0-brane \cite{Frau:1999nyc} in Type~I string theory, which is visible in the K-theoretic analysis of brane charges. Similarly, \cite{Kaidi:2023tqo} argues for the existence of a non-BPS \(\mathbb Z_2\)-valued six-brane in Type~I string theory \cite{Kaidi:2023tqo}, corresponding to the homotopy group \(\pi_1(\operatorname{Spin}(32)/\mathbb Z_2)\) and which trivialises \(\pi_2(\mathrm B(\operatorname{Spin}(32)/\mathbb Z_2))=\pi_1(\operatorname{Spin}(32)/\mathbb Z_2)\).

\subsection{Type~II string theory and M-theory}\label{sec:type-II-m-theory}
Let us briefly discuss what happens for Type~IIA and IIB string theories and M-theory.
In Type~II string theory, for the tower of Ramond--Ramond fields as well as the six-form dual Kalb--Ramond field, the situation is similar: these have modified Bianchi identities \cite[(22.57--58)]{Ortin:2015hya},
so that there are no global symmetries left.
In contrast, at the level of ten-dimensional supergravity, the field strength \(H_3\) of the Kalb--Ramond two-form field is \emph{not} modified, i.e.\ \(\mathrm dH_3=0\) \cite[(22.58)]{Ortin:2015hya}.
Thus, it appears that this gives rise to a global magnetic six-form symmetry,
violating the swampland no-global-symmetries conjecture and \cref{conj:no-generalised-global-symmetry}.
Of course, this is simply because one has neglected objects that couple magnetically to the Kalb--Ramond field, namely the NS5-brane.
That is, smeared NS5-branes give rise to a five-form current \(J_\mathrm{NS5}\) such that
\begin{equation}\label{eq:ns5-current}
    \mathrm dH_3 = J_\mathrm{NS5},
\end{equation}
trivialising the would-be global symmetry. A similar situation arises if we dimensionally oxidise to M-theory in eleven spacetime dimensions. While the seven-form field strength \(G_7\) has a modified Bianchi identity \cite{Fiorenza:2019usl}, at the supergravity level, the Bianchi identity of the four-form field strength \(G_4\) is simply \(\mathrm dG_4=0\).
Again, from the viewpoint of the low-energy supergravity flux sector, this is the correct Bianchi identity. The additional question is what object should classify residual unscreened magnetic defect charges in the full M-theoretic setting, where M5-branes are part of the nonperturbative spectrum. If \(\mathcal A\) is interpreted only as a flux-sector classifier, then the closedness of \(G_4\) is precisely part of the flux equations. If, instead, \(\mathcal A\) is interpreted as a classifier of residual defect charges after dynamically realised brane sectors have been included, then the M5-brane sector should remove the corresponding unscreened magnetic defect charge. In this latter, a smeared M5-brane distribution may be represented effectively by a five-form current \(J_{\mathrm{M5}}\) and one may write schematically
\begin{equation}
    \mathrm dG_4=J_\mathrm{M5},
\end{equation}
which would be the dimensional oxidisation of \eqref{eq:ns5-current}.

More generally, in any consistent theory of quantum gravity, the swampland completeness conjecture \cite{Palti:2019pca} implies that every gauge field strength \(F_p\) should have magnetic sources, i.e.\ \(\mathrm dF_p = J_{p+1}\) for some \(J_p\), which trivialises the would-be magnetic higher-form symmetry whose Noether current would have been \(F_p\).

\paragraph{Hypothesis H.}
Sati and Schreiber have argued, under the name of `Hypothesis H' \cite{Sati:2013rxa,Burton:2018ihd,Fiorenza:2019ain,Fiorenza:2019usl,Sati:2020cml} (for a review, see \cite{Alfonsi:2023pps}), that charges of M-theory should be given by (twisted) four-cohomotopy, which corresponds to taking \(\mathcal A=\mathbb S^4\) to be the four-sphere.

This can be motivated as follows. Eleven-dimensional supergravity has a three-form potential \(C_3\) and a six-form potential \(C_6\); therefore the gauge \(L_\infty\)-algebra is Abelian:
\begin{equation}
    \mathfrak h=\mathbb R[2]\oplus\mathbb R[5].
\end{equation}
On the other hand, if the Pontryagin classes of spacetime vanish,\footnote{In general, there are spacetime-curvature-dependent corrections in the general case, which leads to \emph{twisted} four-cohomotopy \cite{Fiorenza:2019usl}.} the Bianchi identities of the four-form field strength \(F_4\) and the seven-form field strength \(F_7\) are
\begin{align}
    \mathrm dF_4 &= 0, &\mathrm dF_7 &= F_4\wedge F_4.
\end{align}
This means that the adjusted inner-automorphism algebra \(\mathfrak w\) is not the canonical one. That is, the underlying graded vector space of \(\mathfrak w\) is
\begin{equation}
    \mathfrak w = \underbrace{\mathbb R[2]}_{\mathtt c_3}\oplus\underbrace{\mathbb R[3]}_{\mathtt f_4}\oplus\underbrace{\mathbb R[5]}_{\mathtt c_6}\oplus\underbrace{\mathbb R[6]}_{\mathtt f_7}
\end{equation}
whose brackets are given by the Chevalley--Eilenberg algebra
\begin{align}
    \mathrm d\mathtt c_3 &= \mathtt f_4,&
    \mathrm d\mathtt c_6 &= \mathtt f_7-\mathtt c_3\mathtt f_4,&
    \mathrm d\mathtt f_4 &= 0,&
    \mathrm d\mathtt f_7 &= \mathtt f_4^2.    
\end{align}
The adjusted algebra of invariants is therefore
\begin{equation}
    \mathfrak a = \underbrace{\mathbb R[3]}_{\mathtt f_4}\oplus\underbrace{\mathbb R[6]}_{\mathtt f_7},
    \quad\mathrm d\mathtt f_4=0,\quad\mathrm d\mathtt f_7=\mathtt f_4^2.
\end{equation}
This is the Sullivan model of the four-dimensional sphere \(\mathbb S^4\), so that a natural choice of \(\mathcal A\) is simply \(\mathcal A=\mathbb S^4\).
This is not contractible and would produce the would-be higher-form symmetry
\begin{equation}\operatorname H^4(\mathbb S^4;\operatorname U(1))=\operatorname U(1)\end{equation}
of M-theory, corresponding to the Noether current \(G_4\).

However, there is no contradiction between Hypothesis~H and the contractibility expectation discussed above because the two statements concern different interpretations of \(\mathcal A\). Hypothesis~H provides a flux-sector classifier for the \(C\)-field, and in that sense the appearance of \(\mathbb S^4\) is precisely the desired structure. By contrast, the swampland-oriented claims  explained in \cref{ssec:brane} concern a slightly different object: a classifier of residual unscreened defect charges after dynamically realised brane sectors have been taken into account. From this latter viewpoint, one expects the M5-brane sector to remove the residual magnetic defect charge associated with \(G_4\), although the full non-perturbative formulation of this statement would be subtler than the source equation above.\footnote{For recent work on considerations involving M5-branes, see \cite{Giotopoulos:2024sit}.}

\section*{Acknowledgements}
The authors thank Leron Borsten\textsuperscript{\orcidlink{0000-0001-9008-7725}} and Urs Schreiber\textsuperscript{\orcidlink{0000-0002-2876-8877}} for helpful discussion and pointers to the literature.

\newcommand\cyrillic[1]{\foreignlanguage{russian}{#1}}
\newcommand\jafont[1]{\begin{CJK*}{UTF8}{min}#1\end{CJK*}}
\hyphenation{Sza-bo}
\bibliographystyle{unsrturl}
\bibliography{biblio}

\end{document}